\documentclass[12pt,a4paper]{article}
\usepackage{epsfig}
\usepackage{amstex}

\def\nostrocostrutto#1\over#2{\mathrel{\mathop{\kern 0pt \rlap 
  {\raise.2ex\hbox{$#1$}}}
  \lower.9ex\hbox{\kern-.190em $#2$}}}
\def\lsim{\nostrocostrutto < \over \sim}   
\def\gsim{\nostrocostrutto > \over \sim}   

\newcommand{\eref}[1]{(\ref{#1})}      

\newcommand{\epem}{\mbox{$e^+e^-$}}
\newcommand{\pperp}{p_{\perp}}
\newcommand{\Pperp}{P_{\perp}}
\newcommand{\ppar}{p_{\parallel}}
\newcommand{\tg}{{\rm tg}}
\newcommand{\wh}[2]{(\widehat{#1#2})}

\catcode`@=11
\newcount\@tempcntc
\def\@citex[#1]#2{\if@filesw\immediate\write\@auxout{\string\citation{#2}}\fi
  \@tempcnta\z@\@tempcntb\m@ne\def\@citea{}\@cite{\@for\@citeb:=#2\do
    {\@ifundefined
       {b@\@citeb}{\@citeo\@tempcntb\m@ne\@citea\def\@citea{,}{\bf ?}\@warning
       {Citation `\@citeb' on page \thepage \space undefined}}%
    {\setbox\z@\hbox{\global\@tempcntc0\csname b@\@citeb\endcsname\relax}%
     \ifnum\@tempcntc=\z@ \@citeo\@tempcntb\m@ne
       \@citea\def\@citea{,}\hbox{\csname b@\@citeb\endcsname}%
     \else
      \advance\@tempcntb\@ne
      \ifnum\@tempcntb=\@tempcntc
      \else\advance\@tempcntb\m@ne\@citeo
      \@tempcnta\@tempcntc\@tempcntb\@tempcntc\fi\fi}}\@citeo}{#1}}
\def\@citeo{\ifnum\@tempcnta>\@tempcntb\else\@citea\def\@citea{,}%
  \ifnum\@tempcnta=\@tempcntb\the\@tempcnta\else
   {\advance\@tempcnta\@ne\ifnum\@tempcnta=\@tempcntb \else \def\@citea{--}\fi
    \advance\@tempcnta\m@ne\the\@tempcnta\@citea\the\@tempcntb}\fi\fi}
\catcode`@=12

\begin{document}

\setcounter{page}{0}
\thispagestyle{empty}

\noindent
\rightline{CERN-TH/97-199}
\rightline{MPI-PhT/97-68}
\rightline{November 1997}
\vspace{1.0cm}
\begin{center}
{\Large \bf Perturbative Universality in Soft Particle Production} 
\end{center}
\vspace{0.4cm} 
\begin{center}
VALERY A. KHOZE~$^{a, b, }$\footnote{e-mail: khoze@vxcern.cern.ch} \ , \ 
SERGIO LUPIA~$^{c, }$\footnote{e-mail: lupia@mppmu.mpg.de} \ , \ 
WOLFGANG OCHS~$^{c, }$\footnote{e-mail: wwo@mppmu.mpg.de} 
\end{center}

\begin{center}
$^a$ \ {\it INFN - Laboratori Nazionali di Frascati, P.O. Box 13,\\
I-00044 Frascati, Italy}\\ 
\mbox{ }\\
$^b$ \ {\it TH Division CERN\\
CH-1211, Geneva 23}\\
\mbox{ }\\
$^c$ \ {\it Max-Planck-Institut f\"ur Physik \\
(Werner-Heisenberg-Institut) \\
F\"ohringer Ring 6, D-80805 Munich, Germany} 
\end{center}
\vspace{1.0cm}

\begin{abstract}
The spectrum of  partons in a QCD jet becomes independent
of the primary energy in the low momentum limit. This follows within the
perturbative QCD  from  the colour coherence
in soft gluon branching. 
Remarkably, the hadrons follow such behaviour closely,
suggesting the parton hadron duality picture to be appropriate also for 
the low momentum particles. More generally, this scaling property holds for
particles of low transverse and arbitrary 
longitudinal momentum, which explains an
old experimental observation (``fan invariance''). 
Further tests of the perturbatively based picture for soft
particle production  are proposed 
for three-jet events in \epem \ annihilation and di-jet production 
events in $\gamma p$, $\gamma\gamma$
and $p\bar p$ collisions. They are based upon the difference in the  
intensity of the soft radiation from primary $q\bar q$ and $gg$
antennae. 
\end{abstract}

\newpage

\section{Introduction}

The study of multiparticle production in hard processes provides
one with an important information about both the QCD partonic
branching processes and the transition from the coloured partons to
the colourless hadrons. The vast amount of experimental data collected
in various hard collisions processes convincingly confirms that the
inclusive characteristics of QCD jet systems can be successfully
described within the analytical perturbative approach to multiparticle
production, see e.g. \cite{bcm,dkmt1,ko}.
This approach is based on the so-called Modified Leading Logarithmic
Approximation (MLLA) \cite{dt,ahm1} and on the concept of Local Parton Hadron
Duality (LPHD) \cite{adkt1}.

In particular, the experiment clearly demonstrates that the bright colour
coherence effects survive the hadronization stage and are distinctively
visible in the data. This has been known for quite a long time for \epem
annihilation processes (see e.g. \cite{ko}). Recently the very impressive data
from HERA \cite{H10,H1,zeus} and from the TEVATRON 
\cite{cdf,cdfkor} have become available.

The LPHD allows one to relate the (sufficiently) inclusive hadronic
observables to the corresponding quantities computed for the cascading
partonic system.
One of the well known (but still striking) predictions of the perturbative
scenario is the depletion of the soft particle production and the resulting
approximately Gaussian shape of the inclusive distribution in the variable
$\xi  = \ln (P/E)$
for particles with energy $E$ in a jet of energy $P$ (the so-called
``hump-backed plateau'') \cite{adkt1,dfk1,bcmm}.

This behaviour is  
observed in all inclusive energy distributions; the charged particle
spectrum  for not too
small momenta prove to be in surprisingly good agreement with the MLLA-LPHD
predictions reflecting the QCD cascading picture of multiple hadroproduction.
Moreover, the data collected in various hard scattering processes (\epem, DIS,
$\bar p p$) clearly demonstrate a remarkable universality of particle spectra
assuming the proper (MLLA-based) choice of the cascading evolution variables,
equivalent to the \epem $cms$ energy $\sqrt{s}$.

The most challenging is the soft momentum end of the particle spectra
(p~$\lsim1~$GeV) where the non-perturbative dynamics
could invalidate the perturbatively based expectations. An attempt to stretch
the perturbative predictions to the limit of their applicability (or better
to say, beyond it) has been performed in previous papers of the present
authors (see e.g. \cite{lo,klo1,conf,lo2}). 
In these papers it was shown that (after the proper
modifications) the perturbatively based formulae allow a sufficiently smooth
transition into the soft momentum domain.
The non-perturbative hadronization effects are encoded in the 
transverse momentum cutoff $Q_0$ which can be motivated by 
the space-time picture of hadroproduction in QCD jets, see
e.g. \cite{dkmt1}. Explicit predictions for distributions of 
hadrons in the soft region need some additional assumptions on mass effects. 

Scaling phenomena are expected
to be of more general nature.  Let us recall that the gluons
of  long wave length are emitted by
the total colour current which is independent of the internal structure of
the jet and is conserved when the partons split. Applying the
LPHD hypothesis one then expects that the
hadron spectrum at very low momenta $p_h$ should be nearly independent of the
jet energy  \cite{adkt1,lo,vakcar}. 

As discussed in \cite{klo1}, the low momentum data support 
the basic ideas of QCD coherence and LPHD.
Quantitatively, the analysis  was performed in terms of the
invariant particle density $E \frac{dn}{d^3p}$ for \epem annihilation
into hadrons at low momenta in quite a wide $cms$ energy region 
(from ADONE to LEP-2).
The spectra were found to be in a good agreement with the scaling behaviour
and with analytical perturbative expectations which became sensitive to the 
running of the coupling $\alpha_s $ at small scales \cite{lo,klo1}.
 The new H1 data\cite{H1} follow these predictions as well,  
thus confirming the universality of soft particle production.

In this paper we study further the phenomenology of the soft particle
production for both the charged and identified hadrons. The analytical
calculations are compared with the quantity $\frac{dn}{d^3p}$ which has
some practical advantages over the invariant density considered before.
We derive a scaling prediction for particles at low transverse but arbitrary
longitudinal momenta, 
which generalizes our previous results and which for very small 
$\pperp$ explains the so-called
\lq\lq fan-invariance'' experimentally observed long ago~\cite{tassofan}. 

An important test of this line of approach towards soft particle production
is the sensitivity to the colour charge of the primary emitters:
the  density of low momentum particles roughly doubles when going from a
primary $q\bar q$ to a $gg$ antenna. 
It is interesting to note that this prediction applies 
to the soft particles already at
present energies, whereas for the total multiplicity this doubling 
needs much higher
energies.  The sensitivity of the low momentum particle density to the
primary colour charges can be studied in experimentally accessible
processes, such as in $e^+e^- \to 3$ jets \cite{klo1}. Here 
this process is discussed further and   we derive
predictions for the soft particle production in 
quark and gluon exchange reactions which are observed in
photoproduction or DIS at HERA or in $\gamma\gamma$ collisions.

\section{Perturbative predictions for particle production in the soft limit}

In refs. \cite{klo1,conf} an approximate solution of the MLLA evolution
equation for the inclusive energy distribution $D_A^g$ of soft gluons,
originating from a primary parton $A$, has been
 derived.
This solution can be written as
\begin{equation}
D_A^g(\xi,Y) = \frac{C_A}{N_C} D_g^g(\xi,Y)|_{DL} 
     \exp(- G) \quad , \quad 
G = \frac{a}{4N_C} \int_{\xi}^{Y} \gamma_0^2(y) dy
\label{dmlla}
\end{equation}
with
\begin{equation}
D_g^g(\xi,Y)|_{DL}=\beta^2 \ln  \frac{Y-\xi+\lambda}{\lambda} 
 + \beta^4 \int_0^{Y-\xi}
d\tau \ln \frac{\tau+\lambda}{\lambda}
\ln \frac{\tau+\xi+\lambda}{\tau+\lambda}+\cdots
\label{duetermini}
\end{equation}
\noindent
Here we have used the logarithmic variables $\xi = \ln(1/x) = \ln(Q/E)$,
$Y = \ln(Q/Q_0)$ and $\lambda = \ln (Q_0/\Lambda)$
with $E$ the particle energy and $Q$ the jet virtuality $(Q = P\Theta $
for a jet of primary momentum $P$ and half opening angle $\Theta $);
$C_A$ is the respective colour factor, i.e., $C_g = N_C$ and
$C_q = C_F$; $\gamma_0$ denotes the anomalous dimension of multiplicity
and is related to the $QCD$ running coupling by $\gamma_0^2 = 
4 N_C\alpha_s/2\pi$ or $\gamma^2_0 = \beta^2/\ln(\pperp/\Lambda)$ with
$\beta^2 = 4 N_C/b$, $b \equiv (11 N_C - 2n_f)/3$; $\Lambda$ is the
QCD-scale and $N_C$ and $n_f$ are the numbers of colours and of flavours
respectively, $a = \frac{11}{3} N_C + \frac{2n_f}{3N_C^2}$.
The shower evolution is terminated by the transverse
momentum cut-off $\pperp \ge Q_0$.

The first term in (\ref{duetermini}) 
corresponds to the emission of a single gluon and
yields the leading contribution for $E \to Q_0$. 
It is proportional to the colour charge of the primary parton.
Furthermore, this term does not depend on the $cms$ energy, contrary to
the higher order terms which provide the rise of the spectrum for large
$E$ with increasing $\sqrt{s}$. The spectrum vanishes at $E \to Q_0$ as
\begin{equation}
  D_A^g (\xi, Y) \approx \frac{C_A}{N_C} \beta^2 \frac{(E-Q_0)}{\lambda Q_0}
  \label{Dslimit}
\end{equation}

In order to relate  parton and hadron distributions at low momentum $p \sim
Q_0$, one has to make some additional assumptions on how to include mass
effects. In \cite{lo,klo1} it was required 
that the invariant density $E_h\frac{dn}{d^3p_h}$
for hadrons approaches a constant limit when $p_h \to 0$ as observed
experimentally. Here it is worth to recall that at low $p_h$
\begin{equation}
   E_h \frac{dn}{d^3p_h} \sim \frac{\overline{W_1}(s,E_h \sqrt{s})}{s} ,
  \label{ninvar}
\end{equation}
where $\overline{W_i} (s,E_h \sqrt{s})$ are the standard \epem analogues of the
DIS structure functions $W_i(q^2, \nu)$, see e.g. \cite{ikl}.
As well known, 
$\overline W_i (s,E_h\sqrt{s})$ are related to the matrix elements of the
current commutators and should be regular when $p_h \to 0$.

For  spectra which vanish as in (\ref{Dslimit})  a plausible
relation between partons and hadrons can be achieved by relating their
distributions as \cite{dfk1,lo}
\begin{equation}
E_h \frac{dn_A(\xi_E)}{dp_h} = K_h E_p \frac{dn_A(\xi_E)}{dp_p}
    \equiv K_h D_A^g(\xi_E,Y)
    \label{phrel}
\end{equation}
at the same energies
 $E_p=E_h = \sqrt{p^2_h + Q^2_0}  \equiv E \ge Q_0$ where
$\xi_E \equiv \xi = \ln (Q/E)$ and  $K_h$ is a normalization
parameter. If hadrons from both hemispheres are added, $K_h$ should
be replaced by 2$K_h$. For large energies $E_h\gg Q_0$ (\ref{phrel}) 
coincides with the usual LPHD relation $dn/d\xi = K_h D(\xi,Y)$.

In \cite{klo1} we have considered the 
invariant density $E dn/d^3p \equiv dn/dyd^2p_T$ in the
limit of vanishing rapidity $y$ and transverse momentum $\pperp$ 
or, equivalently,
for vanishing momentum $|\vec p| \equiv p$, i.e.,
\begin{equation}
I_0 = \lim_{y \to 0, p_T \to 0} E \frac{dn}{d^3p} = 
\frac{1}{2} \lim_{p \to 0} E \frac{dn}{d^3p}
\label{izero} 
\end{equation}
where the factor 1/2 takes into account that both \epem 
hemispheres are added in the limit $p \to 0$.

Then, since $E_h-Q_0\approx p_h^2/(2Q_0)$ for small $p_h$, 
the invariant hadronic density 
\begin{equation}
E_h \frac{dn}{d^3p_h} = K_h D_A^g(\xi, Y)/(4\pi p^2_h) 
\label{phrel1}
\end{equation}
approaches the finite, energy independent 
limit 
 as in (\ref{izero})
\begin{equation}
I_0=K_h \frac{C_A\beta^2}{8\pi N_C\lambda Q_0^2}.
\label{limit}
\end{equation}
In the fixed $\alpha_s$ limit $\beta^2/\lambda $ is replaced by
$\gamma_0$ and $I_0 \sim 1/Q^2_0$. With prescription (\ref{phrel}) 
one obtains a good description of the
moments of the energy spectrum $D(\xi, Y)$
by the MLLA formulae  with $Q_0$ = 270 MeV in a wide energy range
\cite{lo}, as well as of the particle spectrum in the soft region ($E_h<1$
GeV) \cite{klo1}.

Certainly at the moment there is no unique recipe allowing one to perform
a smooth transition between the perturbative and non-perturbative regimes.
Also there are certain approximations involved in the analytical results
for low energy particles.
In refs. \cite{DKTInt,dkt9} an alternative prescription
has been discussed:
\begin{equation}
   \frac{dn}{d\ln p} = 
    \left(\frac{p}{E}\right)^3 D_A^g(\xi_E, Y)|_{{\rm LS}},
\label{phrel2}
\end{equation}
 where $D_A^g (\xi_E, Y)|_{\rm LS}$ is the so-called
Limiting Spectrum \cite{dt,adkt1} which solves the MLLA evolution equations for
$\lambda$ = 0 in the full kinematic region, except very close to the
boundary $\xi_E$ = Y. At this phase space boundary the Limiting Spectrum
approaches the finite limit \footnote{We are grateful to
Yu.~L.~Dokshitzer for recovering this result.}
$D_g^g(Y,Y)=\frac{4N_C}{a}$, which equals numerically 
1.069~(1.055) for $n_f$=3(5). 

Basing on the prescription (\ref{phrel2})  
one arrives at the alternative formula for the
invariant hadronic density which in the case of \epem annihilation reads
\begin{equation}
  E_h \frac{dn}{d^3 p_h} = 2K_h\left(\frac{1}{4\pi E_h^2}\right)
   D_q^g(\xi_E, Y)|_{{\rm LS}},
\label{phrelLS}
\end{equation}
with $ D_q^g= \frac{C_F}{N_C} D_g^g$.
As it is easy to see this prescription also guarantees that the hadronic
density reaches a constant limit at $p_h \to 0$.
We shall compare here the expectation based on Eq.~(\ref{phrelLS}) with the
experimental data for charged pions $(K_h = K_{\pi})$. The MLLA Limiting 
Spectrum taken at $\Lambda = Q_0 $ = 150 MeV has been found to provide 
a good description of pion spectra at relativistic energies, see e.g.
\cite{adkt1,dkt9}. 
Recall that the low momentum region in charged particle 
spectra is dominated by the pions.

\section{Phenomenology of soft particle production}

The energy dependence of the soft particle production rate was
investigated previously for \epem annihilation 
in terms of the invariant particle density
$E dn/d^3p$ in the wide energy interval (from ADONE to LEP-2) \cite{klo1}.
Recently, new data on soft particle production became available from DIS at
HERA \cite{H1}. In the following we compare various results for the
quantity $dn/d^3p$ as a function of the momentum $p$ of the registered 
particle. When presented in terms of this observable the data points are 
independent of the choice of the theoretical parameter 
 $Q_0$; in this case, however, we are dealing with a quantity which is not
directly related to the Lorentz-invariant phase space.
Using this quantity may facilitate
 the comparison between
different experimental results.

\subsection{Energy dependence of soft particle production}
 
 Fig.~1 shows the momentum dependence of  $dn/d^3p$ 
for various $cms$ energies $\sqrt{s}$ in  \epem  annihilation 
in comparison with the theoretical predictions. In Fig.~1a the 
distributions of charged particles 
are presented in comparison with the approximate MLLA result
(\ref{dmlla}) with prescription (\ref{phrel}) for the relation between
partons and hadrons. The theoretical prediction describes well the 
approach to the scaling limit for low  momenta and also the shape and
energy dependence of the spectra with the choice $Q_0=270$ MeV, as already
discussed in \cite{klo1}. The detailed behaviour at the very small momenta
$p\lsim Q_0$ depends on the approximation at the parton level involving
large production angles ($\vartheta>Q_0/p$)
and also on the assumed relation between partons and
hadrons in this region. 

This point is illustrated in Fig.~1b, where we show
the theoretical predictions with the alternative choice of the low momentum
behaviour discussed in Sect. 2,
based on the approximate analytical MLLA solution
for $Q_0=\Lambda$ (Limiting Spectrum) and relation (\ref{phrel2});
this form describes the pion spectra for $\sqrt{s}\gsim 10$ GeV with the mass
parameter $Q_0=150$  MeV which is  in agreement with the earlier fits to the
pion spectra  
\cite{adkt1,dkt9}.
  With this prescription the 
particle production rate in the soft limit $p\to 0$ 
is lower than in Fig. 1a, but it also approaches an energy
independent value in this limit and that is the essential prediction
of the QCD approach based on  coherent soft gluon emission.

Looking at the data in Fig. 1 more closely, there seems to be a certain
increase of the soft particle rate at LEP energies. This is illustrated in
Fig.~2 a and b where the data below and above 
$\sqrt{s}=91$ GeV respectively in \epem
annihilation show separately a very good approach to an energy independent
limit at small $p$, whereas there is a difference in absolute rates of  
20-30\% in the two energy regions. This is also visible in Fig. 3a where we 
show the $cms$ energy dependence of three values of the momentum $p$.
We also include here the recent data from HERA \cite{H1} which closely match
the \epem results. At the lowest momentum $p=0.2$ GeV we see an almost
energy independent production rate, but a small jump at the LEP-1 energy is
visible. A rise at the higher energies could partially be caused by 
the weak decay products of heavy quark particles 
(perhaps also particles with strange quarks)
which would
add incoherently to the spectrum of particles produced directly from the
soft gluons. This could happen in particular at the $Z^0$ because of the
larger branching ratio into $b$-quarks. However, the experiment 
\cite{delphi0} clearly shows that at low momenta ($p\leq 0.5$ GeV) the
effect induced by the $b\bar b$ events is below 5\%.
It would be interesting to clarify quantitatively 
in more details which
part of the observed rise could be explained by the overall effect of the
weak decays.
 A certain small increase of the rate is actually also expected from the 
theory at the nonzero momenta (see the curves in Fig. 3a).
For larger momenta the rise of the production rate with energy is more
pronounced and this trend is reproduced by the theoretical calculation
for the full range of energy where the data are available ($\sqrt{s}\gsim 3$
GeV). 

An approximate energy independence of the rate for the low momentum
 kaons and protons  is well tested for
$\sqrt{s}\gsim 10$ GeV where data at some common low momentum value are
available. A comparison in Figs. 3a and 3b suggests that the 
spectra become practically  energy
independent when the particle momenta are
of the order of their  mass or smaller.

It should be noted that the energy independence of the soft particle rate is
by no means trivial. We recall, in particular,
that the particle density $dn/dy$ at central
rapidity $y\approx 0$ rises with $cms$ energy. This is illustrated in Fig. 4
for TASSO and OPAL data where rather precise 
results are
available at $p=0.25$ GeV.
Whereas in this energy range (14 to 91 GeV) the plateau height
rises by about 70\% the corresponding rise for the soft particles at $p=0.25$
GeV is only about 10\%. This is consistent with the QCD picture of particle
production: the central plateau rises asymptotically with the same exponent $\sim \sqrt{\ln
s}$ as the full particle multiplicity and 
this rise comes mainly from the high
transverse momentum particles. The  soft particle production is
suppressed because of the coherence in the emission from all colour
sources.

\subsection{Universal soft limit for different particle species?}

In Fig. 5 we consider for different identified particles with mass $m$ 
the rescaled spectra $m^3 dn/d^3p$ which are expected to approach energy
independent dimensionless numbers in the soft limit $p\to 0$. First we see
again, that in the pion spectra the violation of energy independence occurs
at lower momenta than in the spectra of 
the heavier particles. Furthermore, these rescaled spectra of pions and
kaons, separated by an order of magnitude around $p\sim 0.8$ GeV seem to
converge towards a common value in the soft limit. A similar convergence
is also suggested for the protons although the lack of precise data at small
momenta exclude a firm conclusion. We find 
an approximate universality of the quantity
\begin{equation}
I_0^m=m^3\frac{dn}{d^3p}|_{p\to 0} \approx 0.1\div 0.2    \label{mlimit}
\end{equation}
for the particle species considered here. This 
corresponds to our previous observation \cite{klo1} that the invariant
density behaves like 
$I_0=Edn/d^3p|_{p\to 0}\sim m^{-2}$ in the soft limit. 
In our theoretical picture this quantity is given by 
$I_0\sim 1/Q_0^2$ (and not $1/\Lambda^2$, for example). To the extent that $Q_0$ is
related to the particle mass such a universality could be anticipated. 
It would be
interesting to have more experimental information on low momentum particles
to test better the validity of such a universal behaviour; new  
HERA data with identified particles at such small momenta could be helpful
as well. 

\section{Scaling properties 
of particle production at small $p_\perp$}
\subsection{Colour coherence and \lq\lq fan invariance"}

As discussed before the gluon spectrum in the soft limit becomes
independent of  energy of the primary parton above about 1 GeV.
This energy independence can be observed, for example, in the $cms$ of \epem 
annihilation or in the Breit frame of DIS. This result can be
easily generalized. We make a Lorentz transformation along the initial
parton direction, then in the new frame the initial partons have different
energies but the soft particle production rate 
should remain unaltered as long as the primary energies are still
sufficiently large (and the colour of the primary parton has not changed).
 Consequently, the quantity
\begin{equation}
I_0(y)= \lim_{\pperp \to 0} E \frac{dn}{d^3p}= \lim_{\pperp \to 0} 
E \frac{dn}{dy d^2 \pperp}
\label{izeroy}
\end{equation}
should be independent of rapidity $y$ for such transformations \cite{klo1}
and this behaviour should still hold approximately for $\pperp$
not much larger than the hadron masses. Therefore, if the coherence argument 
is applied to 
this class of frames one obtains the prediction of scaling for the
distribution
\begin{equation}
 \frac{d^2n}{dy d \pperp}=f(y,\pperp).
\label{nykt}
\end{equation}
Equivalently, for $y\approx \eta=-\ln\tg (\vartheta/2)$ with production angle
$\vartheta$ one expects
\begin{equation}
 \frac{d^2n}{d \ppar d\vartheta}=f(\ppar,\vartheta).
\label{fan}  
\end{equation}            
An approximate scaling behaviour of this type -- but for the distributions
normalized to unity -- was found by the TASSO collaboration
\cite{tassofan} in the energy range from 14 to 34 GeV long ago and was
called \lq\lq fan invariance''.\footnote{We thank P. M\"attig for pointing
out to us this scaling law which triggered the study in this
section.}
In the following we will show explicitly how this scaling behaviour arises
in the perturbative QCD calculations; furthermore we predict how the
abovementioned scaling behaviour is violated.
This will lead to the
suggestion of other variables in which scaling is restored in a wider
kinematical region.

\subsection{Scaling properties of the analytical results}

The double differential distribution $dn/d\xi d\vartheta$ in $\xi=\ln(1/x)$
and angle $\vartheta$ is derived from the $\xi$-distribution of particles 
$D(\xi,Y)$ in a jet of half opening angle $\Theta$ by  (see also \cite{dfk1})
\begin{equation}
\frac{d^2n}{d\xi d\ln\vartheta} = \frac{d}{d\ln \vartheta}
D_A^g(\xi,Y)\vert_{\Theta=\vartheta}.  \label{dndtheta}
\end{equation}
with $Y=\ln (P\Theta/Q_0)$ for small angles.
For larger angles an appropriate variable is $\tilde
Y=\ln (2P\sin(\Theta/2)/Q_0)$ instead of $Y$ \cite{do}. For the 
application at small $\pperp$ we take the 
approximate expression~(\ref{dmlla})  for $D_A^g(\xi,Y)$ and obtain
(with $\pperp=E\vartheta$)
\begin{equation}
\frac{d^2n}{d\xi d\ln\pperp} = \frac{d}{d Y}
D_A^g(\xi,Y)\vert_{DL}e^{-G} - \frac{\partial G}{\partial Y}
    D_A^g(\xi,Y).  \label{dnMLLA}
\end{equation}
Within the DLA this distribution 
(the first factor in the first term) is given by 
\begin{equation}
\frac{d^2n}{d\xi d\ln\pperp} = \frac{C_A \beta^2}{N_C}\left[
\frac{1}{\ln\frac{\pperp}{\Lambda}}
+\beta^2\ln\left(\frac{\ln\frac{\pperp}{\Lambda}}
{\ln\frac{Q_0}{\Lambda}}\right)
\ln\left(\frac{\ln\frac{\Pperp}{\Lambda}}
{\ln\frac{\pperp}{\Lambda}}\right)\right]
 \label{dnDLA}
\end{equation}
\begin{equation}
Y-\xi+\lambda=\ln\frac{\pperp}{\Lambda}, \qquad 
Y+\lambda=\ln\frac{\Pperp}{\Lambda}=\ln\frac{\pperp/x}{\Lambda}.
\label{notat} 
\end{equation}

For small $\pperp$ the first term in the brackets from the single gluon
bremsstrahlung dominates and is energy independent. This is the term which
corresponds to the \lq\lq flat plateau'' of radiation in the low $\pperp$
limit; it is  also  responsible for the \lq\lq fan invariance'' as follows
from the exact relation 
\begin{equation}
\frac{d^2n}{dp_\parallel d\vartheta}=\frac{d^2n}{d\xi_p d\pperp}
  \label{fandef}
\end{equation}
with $\xi_p=\ln (P/p)$ defined in terms of the  momentum $p$.
The second term gives the energy dependent correction which vanishes in the
soft limit $\pperp \to Q_0$; the $cms$ energy $\sqrt{s}$ 
 enters only through the variable
$\Pperp=\pperp/x$ with $x=2E/\sqrt{s}$.
The MLLA correction involves the exponent $G$ and its derivative 
\begin{equation} 
G(\xi,Y)=\frac{a}{b} \ln\left(\frac{Y+\lambda}{\xi+\lambda}\right) 
\quad , \quad 
\frac{\partial}{\partial Y}G(\xi,Y)=\frac{a}{b(Y+\lambda)} \quad , 
\end{equation} 
so it again
depends only on $Y+\lambda$ and $\xi$ (for soft particles one can assume
$\xi\gg\lambda$). Therefore our perturbative results fulfil the scaling law
\begin{equation}
\frac{d^2n}{d\xi d\ln \pperp} = F(\xi,\pperp),
\label{Ffun}
\end{equation}
i.e. the dependence on the $cms$ energy enters only through the variable 
$\xi$ or $x$. In this way the violation of the 
\lq\lq fan invariance'' can be partly absorbed into the scaling variables
$\xi$ or $x$. 

The analytical predictions have been derived for the soft gluon radiation 
(with energies smaller than, say, ${\cal O}$(1 GeV)),
while the angles are measured with respect to the primary parton axis.
There are two effects which are beyond the perturbative
calculations. 

First, the direction of the primary parton is not directly known,
but may be related to the jet axis or to the initial hadron or
photon direction. Here the situation can be improved by measuring the
energy-multiplicity correlation \cite{dkmt1} taking the momentum of each
particle in the event in turn as jet axis with the particle energy fraction 
as a weight. The angular correlation of this type has been studied using Monte
Carlo events with favourable results \cite{ow}. 

Secondly, we can only 
perform calculations  at the parton level, so in comparison
with experiment an assumption on the effect of hadronization has to be made. 
This will be described in the next subsection.

\subsection{Relation between parton and hadron distributions}

According to the LPHD concept, we assume that 
parton and hadron spectra are proportional to each other 
if momenta are large compared to $Q_0$.  
This picture has so far only been applied to 
observables which are integrated over $p_\perp$, while we are now interested
in distributions differential in $p_\perp$. Since perturbative
formulae explicitly contain the cut-off $p_\perp>Q_0$, they cannot indeed be 
directly compared with the data at small $p_\perp$, but they have to be 
modified in order to provide a smooth behaviour in this region. 
There is no unique prescription 
to be followed in order to build this bridge between parton and hadron level, 
and what we are proposing here is  a simple phenomenological ansatz 
which is consistent with kinematical requirements and the scaling behaviour
in the soft limit. 
 
In analogy to the relation (\ref{phrel}) applied to the energy spectra in
Section 2, we take the transverse momentum cut-off $Q_0$ as an
effective mass of the hadron with $E_h^2=p_h^2+Q_0^2$
and we then replace
\begin{equation}
p_{\perp,p}^2=p_{\perp,h}^2+Q_0^2,
 \qquad  E_p = E_h 
   \label{htop}
\end{equation}
in eqs.~(\ref{dnMLLA}) and (\ref{dnDLA})
 on the $r.h.s.$ as well as on the $l.h.s.$ This choice
guarantees the proper kinematical limits  $p_{\perp,p} \ge Q_0$ 
 and $p_{\perp,h} \ge 0$ for parton and hadron spectra, while they coincide
for large $p_\perp$. 
Furthermore we use as before  
$E\frac{dn}{dp}\vert_h=\frac{dn}{d\xi_E}\vert_p$. So for hadrons one can
write (dropping the label $h$ everywhere)
\begin{equation}
E\frac{dn}{dp
d\pperp}=\frac{\pperp}{\pperp^2+Q_0^2}F(\xi_E,\sqrt{\pperp^2+Q_0^2})
\label{edndpkt}
\end{equation}
with the function $F$ defined by (\ref{Ffun}) at the parton level,
whereby the replacements (\ref{htop}) are applied. 
Once again, relations such as (\ref{htop}) 
are not an essential part of the perturbative
LPHD picture, rather they should be considered as a plausible extension into
the region $\pperp\lsim Q_0$, consistent with kinematics and 
scaling properties.

Combining now (\ref{edndpkt}) and (\ref{fandef}), one obtains  
quantitative theoretical predictions for the angular distribution of
particles. Results for different values of $p_{\parallel}$ at 
$\sqrt{s}$ = 14  GeV are shown in Fig.~\ref{figfan} 
for $Q_0$ = 0.5 GeV and $\lambda = 0.05$.  
Choosing such parameters, the results 
describe reasonably well the experimental 
data obtained  by the TASSO collaboration\cite{tassofan} in the region of low
$p_{\parallel}$, while some deviation is visible for larger values of
$p_{\parallel}$\footnote{This value for $Q_0$ is largely determined by the
position of the maximum in the angular distribution and is larger than the
value taken above for the energy spectrum. A unification is possible in
principle by calculating the energy spectrum from the integration over the
modified double differential spectrum $d^2n/dEdp_\perp$, but this is not
attempted here.}. The large values of  $p_\parallel$ and of angles
correspond, however, 
 to large values of the transverse momentum $p_\perp$, where our 
approximations are not valid any more. 
This point is particularly
important for the study of ``fan invariance'', i.e., 
the energy independence of the renormalized angular 
distribution: the distributions presented by the 
TASSO collaboration\cite{tassofan} are
normalized to unity in the full angular region and therefore not directly
accessible by our calculations. 

A better test of the scaling law~\eref{fan} can be obtained from 
 the direct study of the double differential distribution
$d^2n/dp_\parallel dp_\perp$ in absolute normalization (i.e., normalized to
the particle multiplicity) as a function of the transverse momentum $p_\perp$
at fixed $p_\parallel$. 
Fig.~\ref{energy} shows the theoretical predictions in 
MLLA with the same parameters as in Fig.~\ref{figfan}  
 at $p_\parallel$ = 0.8 GeV at four different values 
of $cms$ energy $\sqrt{s}$.  
A good scaling behaviour is visible in the region of very low transverse
momentum, whereas the position of the peak and the tail are found to
smoothly depend on the $cms$ energy. 
If the distributions were rescaled to the same area as in \cite{tassofan},
the overall energy dependence would be rather small. 

In order to better understand the origin of this energy 
dependence, it is useful to separately look at the contribution of 
the different terms in the theoretical formula. This is shown in
Fig.~\ref{approx}, where the $d^2n/dp_{\parallel} dp_\perp$ distributions  
at $p_\parallel$ = 0.8 GeV and $\sqrt{s}$ = 34 GeV are shown for different 
approximations:  
the DLA (Born term only), DLA (Born plus next order term) and  MLLA.   
Whereas the Born term of DLA, which is 
independent of $cms$ energy, is dominant at small $p_\perp$, 
the other terms become important with increasing $p_\perp$ and give rise to 
the energy dependence visible in Fig.~\ref{energy}. As the next order term
becomes important for larger $p_\perp$, we also compare in
Fig.~\ref{approx} with the prediction from the Limiting Spectrum  
which corresponds to an all order summation within a high energy
approximation (taking  the same $Q_0$,  $\lambda$ = 0 and normalization 
($K_h$ = 3.5) adjusted to match the MLLA result).  
It is remarkable  how close are the predictions from the MLLA 
and  from  the Limiting Spectrum. This
result suggests that retaining only the first two terms in the DLA
solution gives indeed a reasonable approximation  
in the region of low transverse momentum. 

The deviation from the scaling behaviour~\eref{fan} is due to the weak
logarithmic dependence of the correction terms on the primary energy. 
As a consequence of~\eref{edndpkt} with $p\sim p_\parallel$ the invariant
density, after our modification~\eref{htop}, still follows the scaling
behaviour 
\begin{equation} 
E\frac{dn}{d^3p} = f(x_E, p_\perp) \quad , 
\label{neweq}
\end{equation}
i.e.,  for fixed $p_\perp$ it depends on particle and $cms$ energies only
through $x_E = 2E/\sqrt{s}$. 
This scaling behaviour~\eref{neweq} is expected to hold in a wider kinematic
region than eq.~\eref{fan}.

\section{Further tests of the perturbatively-based  
picture of soft particle production}
\subsection{Sensitivity to the colour charge of primary partons}
The nearly energy independent soft end of the particle spectrum has been
interpreted above as a consequence of the coherent soft gluon emission
together with the LPHD. Alternatively, one may argue that at these low
momenta  of few hundred MeV we observe the purely hadronic phenomena not related to
the perturbative QCD at all. It is therefore important to test further the 
perturbative nature of the soft particle production.

The production rate in the limit (\ref{izero}) cannot be predicted in
absolute terms, but we note the dependence of this rate on the colour charge
factors $C_F$ and $N_C$ for the jets originating from a primary quark
or gluon respectively. Therefore, we have argued in \cite{klo1} that the
sensitivity of the soft end of the spectrum to the colour of the primarily
produced hard partons provides a crucial test of the perturbatively based
picture and can teach us about the  region of its validity.

The prediction for  the ratio of multiplicities in gluon and quark
jets  $R_{g/q}=N_C/C_F=2.25$  has been made long ago \cite{bg} but holds
only at asymptotic energies. At existing energies this ratio is measured 
to be smaller but not in disagreement with the perturbative QCD analysis
(see, for example \cite{ko,opalglu,lo2}). Our present discussion concerns only the
soft particles: in this kinematic regime the asymptotic predictions 
 on the spectrum, in particular the scaling properties, are well
satisfied and therefore the asymptotic ratio $R_{g/q}$ in this limit could
be asymptotic too without contradicting the global results. 

Therefore, in order to test this scenario, one would like to compare the soft
particle production in  $q\bar q$ and $gg$ events. Whereas there are well
measured spectra down to small momenta of about 200 MeV
from quark jets in \epem and DIS processes (see Figs. 1 and 2)
the direct production of a $gg$ final state is more difficult.\footnote{At
high energy \epem colliders the process $\gamma\gamma\to gg$ may become a
 prospective source of clear information \cite{vak2}}  
To perform such a comparison nevertheless we have 
discussed in \cite{klo1} two different approaches.

First, there is the comparatative study of two jet events 
arising from  reactions with  primary
hadrons and photons mediated by quark or gluon exchange processes. 
If the exchange is sufficiently hard the two types of processes yield
primary colour sources of the desired  $q\bar q$ and $gg$ types.
We will dicuss below how these limits can be derived explicitely from
the hard scatterings in the limit of small momentum transfer.
The phenomenological studies along this line are described in papers
\cite{klo1}.

Secondly, one may consider final states with more than two partons.
In certain collinear limits such final states approach a 
$q\bar q$ or $gg$ type antennae. A simple example is provided by the process
$\epem\to q\bar q g$ which for the configurations with angle 
$\Theta_{qg}\sim 0$ or $\Theta_{\bar qg}\sim 0$ behaves like a $q\bar q$
and for  $\Theta_{\bar qq}\sim 0$ behaves like a  $gg$ final state 
\cite{gary}. Such extreme configurations are in general not realistic.
However, we can study the radiation perpendicular to the event plane (say, in
the cone with not too large opening angle $\delta$), then the  particle density
at low momenta should vary by the factor $N_C/C_F$ when going from the
$q\bar q$ to the  $gg$ type configuration and this variation can be
predicted also for the intermediate kinematical range.
Similarly, one can study such an associated perpendicular radiation in 
parton-parton scattering processes with quark or gluon exchange at any angle
(see Fig.~9). 

In the following we will present the expectations for   
various processes; the main results for $\epem\to q\bar q g$ 
have been given already in \cite{klo1}.  

\subsection{Soft radiation in $\epem\to q\bar q g$}

It was realized long ago that the overall
structure of soft particle angular distribution in multijet events in hard
processes is governed by the underlying dynamics of colour at small
distances. 
An instructive known example is the perturbative explanation \cite{adkt} of 
the particle angular flow in  $e^+e^-\to q \bar q g$ events (\lq\lq
string effect'' \cite{lund}). The multiplicity flows build up a colour
portrait of an event which can
be used as a natural \lq\lq
partonometer'' mapping the primary interaction short-distance process 
\cite{emw,dkt8,dkmt1}. 
In these studies the drag effect on the particles within the production plane
 is caused by the primary colour dipoles. More recently, the azimuthal angular
distribution around the jet directions \cite{eks,ks}
have been studied. The partonometry ideas are strongly
supported by a wealth of experimental data 
 from $e^+e^-\to q \bar q g$ annihilation (see e.g. \cite{ko})
and the TEVATRON $p \bar p$ collider \cite{cdf}.

In the present application we are interested in the particle production at
low momenta $p$ of order $Q_0$. A straightforward calculation is possible for the 
radiation perpendicular to the production plane of the jets.
Otherwise, in multijet events, the cut-off condition $p_\perp>Q_0$ is more
difficult to implement. Also, in this case $p_\perp$ appears as argument
 of $\alpha_s$ and does not change for boosts of partons
within the plane. In the collinear limits for the 3-jet events one
obtains the  radiation transverse to the jet for which the coherence
arguments apply in the first place.

In the simplest case of a boosted $q\bar q$ pair, or equivalently, in 
 $\epem \to q\bar q \gamma$
the radiation pattern of gluons with momentum $p$ in direction
$\vec n$ is given by \cite{adkt}
\begin{eqnarray}
\frac{dN_{q\bar q}}{d\Omega_{\vec n}dp} 
  & =& \frac{\alpha_s }{(2\pi)^2p} W^{q\bar q}(\vec n)
\label{Nqqbg}\\
W^{q\bar q}(\vec n)&=& 2C_F \wh{i}{j},\qquad 
\wh{i}{j}=\frac{a_{ij}}{a_ia_j}  \label{ijdef} 
\end{eqnarray}
where  $a_{ij} = (1 - \vec{n}_i\vec{n}_j)$ 
and $a_i = (1 - \vec{n}   \vec{n}_i)$ 
for $q$ and $\bar q$ in direction $\vec n_i$ and  $\vec n_j$. 
For the radiation perpendicular to the plane we simply obtain 
\begin{equation}
\wh{i}{j}=1-\cos\Theta_{ij} \label{whperp}
\end{equation}
with the relative angle $\Theta_{ij}$ between the primary partons. This yields
the two limits $\wh{i}{j}=0,2$ for parallel and antiparallel momenta
respectively.

As we are interested only in the soft particles and the relative variations
between different processes or configurations we restrict ourselves to 
the single gluon bremsstrahlung of order $\alpha_s$ as in (\ref{Nqqbg}).
The MLLA corrections will become important for larger momenta 
(see e.g. the discussion in \cite{adkt,klo1}).
 

The soft gluon radiation in the $q\bar q g$ process is given by the
antenna pattern (defined as in (\ref{Nqqbg}))
\begin{equation}
W^{q\bar q g}(\vec{n}) = N_C [ (\widehat{1+}) + (\widehat{1-})  
- \frac{1}{N_C^2}  (\widehat{+-}) ] \label{wqqbg}
\end{equation}
where $(+,-,1)$ refer to $(q,\bar q,g)$ and this 
bremsstrahlung pattern holds, irrespective
of whether the partons are incoming or outgoing. 
With (\ref{whperp}) one can obtain the particle density in the direction
perpendicular to the production plane.

It is convenient to normalize the production rate of the soft particles to
the corresponding rate in the standard process $\epem \to q\bar q$
in its rest frame with $W^{q\bar q}_\perp = 4C_F$ according to 
(\ref{ijdef}) and define the ratio 
\begin{equation}
R^p_{\perp}(p) \equiv \frac{dN_{\perp}^{p}/d\Omega_{\vec n}dp}
     {dN_{\perp}^{q\bar q}/d\Omega_{\vec n}dp}=\frac{W^p_\perp}{4C_F}.
\label{rperpdef}
\end{equation} 
for a general process $p$. 
For the perpendicular radiation in the 3 jet events one finds
\begin{equation}                                                            
R^{q\bar q g}_{\perp}(p) =
\frac{N_C}{4 C_F} [ 2 - \cos \Theta_{1+} - \cos \Theta_{1-} -
\frac{1}{N_C^2}
(1 - \cos \Theta_{+-} ) ].
\label{rperpqqg}
\end{equation}
In our approximation the momentum spectrum does not depend on the 
angles $\Theta_{ij}$ between the jets. It will be interesting to 
study the spectrum $dn/d^2p$ in such a cone experimentally and to 
find out 
whether the angular dependence of the particle density in (\ref{rperpqqg})
 holds down to small momenta.

It is also interesting to note the difference of this prediction to the
large
$N_C$ approximation in which the $q\bar q g$ event is treated as a  
superposition of two $q\bar q$ dipoles (see, e.g., \cite{lund}). 
In this case the last term in eq.~\eref{rperpqqg} drops out
and $C_F = (N_C^2 - 1)/2 N_C \simeq N_C/2$. This yields
\begin{equation}
R_{\perp} \simeq 
\frac{1}{2} [ 2 - \cos \Theta_{1+} - \cos \Theta_{1-} ] \qquad \hbox{(large
$N_C$)}
\label{rperpln}
\end{equation}
Predictions from these formulae for $R_{\perp}$ are presented in
Table~\ref{tableperp} for various relative angles $\Theta_{ij}$. Note, in 
particular, the limiting case 
$R_{\perp}$ = 1 for soft or collinear primary gluon                        
emission and the proper $gg$ limit for the parallel $q \bar q$             
($\Theta_{+-}$ = 0) configuration, as expected.                            
 
The role of the large-$N_C$ limit
can be investigated also by studying  the production rate in
3-jet events normalized
to the sum of rates from the corresponding 2-jet events (dipoles)
with opening angle
$\Theta_{1+}$ and $\Theta_{1-}$ respectively:
\begin{eqnarray}
\label{rtilde}  
\tilde R_{\perp} &\equiv& \frac{dN_{\perp}^{q\bar q g}}{dN_{\perp}^{q\bar
q}(\Theta_{1+}) + dN_{\perp}^{q\bar q}(\Theta_{1-})} \\
&=& \frac{N_C^2}{N_C^2 -1} \Bigl( 1 - \frac{1}{N_C^2} \frac{1 - \cos
(\Theta_{1+} + \Theta_{1-} )}{2 - \cos \Theta_{1+} - \cos \Theta_{1-}}
\Bigr)
\nonumber
\end{eqnarray}  
This ratio measures directly the deviation from the large-$N_C$ limit
$\tilde R_{\perp}$= 1 and
thereby from the $q\bar q$-dipole approximation. This approximation is not
necessarily limited towards soft particle production. For the simple case of
Mercedes-like events ($\Theta_{1+} = \Theta_{1-} = \Theta_{+-}$) one obtains
$\tilde R_{\perp} = 17/16 = 1.06$.
The 2-jet rates for relative angle $\Theta_{ij}$ which appear in the
denominator of eq.~\eref{rtilde} could be found experimentally from the
corresponding $q\bar q \gamma$ final states.

\begin{table}     
 \begin{center} 
 \vspace{4mm}   
 \begin{tabular}{||c|c|c||}
  \hline
 & $R_{\perp}$ & $R_{\perp}$ (large $N_C$) \\
  \hline
 $\Theta_{1+} = \pi - \Theta_{1-}$ & 1 & 1 \\
 (collinear or soft gluons) & & \\
 $\Theta_{1+} =  \Theta_{+-} = \frac{5}{6} \pi$ &
1.21 &
1.18 \\
 $\Theta_{1+} = \Theta_{+-} = \frac{3}{4} \pi$ &
 1.42 &
 1.35 \\
 $\Theta_{1+} = \Theta_{1-} = \frac{2}{3} \pi$ &
 1.59  &
  1.5 \\
 (Mercedes) & & \\
 $\Theta_{1+} = \Theta_{1-} = \pi$ & $\frac{N_C}{C_F}$ = 2.25 & 2 \\
 ($q\bar q$ antiparallel to $g$) & & \\
 \hline
 \end{tabular}  
 \end{center}   
\caption{Prediction for the ratio $R_{\perp} = dN_{\perp}^{q\bar q
g}/dN^{q\bar
q}_{\perp}$ from \protect\eref{rperpqqg}
and its large-$N_C$-approximation~\eref{rperpln} for different
configurations
of the $q\bar q g$ events ($\Theta_{1+} \equiv \Theta_{gq}$, $\Theta_{1-}
\equiv
\Theta_{g\bar q}$,
$\Theta_{+-} = 2 \pi - \Theta_{1+} - \Theta_{1-}$).}
\label{tableperp}
\end{table}

Let us list a few further results:

\noindent {\it a)}
A particularly simple situation is met for Mercedes-type events
where no jet identification is necessary for the above measurements.

\noindent {\it b)}
The large angle radiation is independent of the mass of the quark (for
$\Theta \gg m_Q/ E_{jet}$).

\noindent {\it c)}
The above predictions for the ratios $R_{\perp},\tilde R_{\perp}$ are
derived for the soft particles according to the bremsstrahlung
formula~\eref{Nqqbg}. One may also consider the particle
flow integrated over momentum, as in the discussion of the string/drag
effect. In this case one has to 
include all higher order contributions which take into account the fact that
the soft gluon is part of a jet generated from a primary parton. Then the
angular flow $dN/d\Omega_{\vec{n}}$ is given by the
product of the radiation factor
$W(\vec{n})$ and a \lq\lq cascading factor''\cite{adkt}. For the ratio of
multiplicity flows
one obtains the same predictions~\eref{rperpqqg} and~\eref{rtilde}
as the cascading factors cancel.
It will be interesting to see to what extent the predicted angular
dependence
for both quantities -- the multiplicity flows and the soft particle yields
--
are satisfied experimentally.

\noindent {\it d)}
The similarity of particle flows in 3-jet and 2-jet events with
corresponding
angles $\Theta_{1+}$, $\Theta_{1-}$, as expected in the large-$N_C$
approximation, should also apply to further details of the final state such
as
the particle ratios. Since the Lorentz transformation along the boost
direction
produces a larger  drag for heavier particles, one may expect
that after the boost the $K/\pi$ and $p/\pi$ ratios for soft
particles decrease and then the same is true for soft particles in 3-jet
events in comparison to 2-jet events in their rest frame\footnote{This
expectation has been verified in the JETSET Monte Carlo\cite{jetset}.
We thank T. Sj\"ostrand for
providing us with this information.}.

\subsection{Soft radiation accompagnying 
photo-production of di-jets}
\subsubsection{Direct and resolved processes}

Next we consider the photoproduction of di-jets. 
The soft brems\-strahlung depends only on the momentum vectors of the partons
but not on the virtuality of the photon. 
Our results can be applied to $\gamma p$ as well as to $\gamma\gamma$
collisions. The di-jet photoproduction has been studied recently in 
detail at HERA \cite{dijzeus,dijh1}. In di-jet production one can distinguish
in the leading order QCD approach the direct and the resolved processes 
 \cite{butter}.
In the first case the photon participates directly in the hard scattering
subprocess and transfers a large fraction ($x_\gamma\sim1$) of its primary
energy to the secondary jets; in the second case, the hard scattering
subprocess involves the partons in the protons and also in the photons
and the energy fraction is 
smaller ($x_\gamma < 1$). The data indeed show two peaks in
the distribution of $x_\gamma$ corresponding to the two types of processes.

The direct processes are mediated by
\begin{eqnarray}
a) \qquad \gamma(p_1)+g(p_2)& \to &q(p_3)+\bar q(p_4) \label{gamgl}\\
b) \qquad \gamma(p_1)+q(p_2)& \to &q(p_3)+ g(p_4), \label{gamq} 
\end{eqnarray}
the so-called photon-gluon-fusion and QCD-Compton processes respectively.
Their differential cross sections are given by
\begin{eqnarray}
\frac{d\sigma^a}{dt}&=& \frac{\pi\alpha\alpha_sQ^2_q}{s^2}\left(\frac{u}{t}+
\frac{t}{u}\right)\label{siga}\\
\frac{d\sigma^b}{dt}&=& \frac{8\pi\alpha\alpha_sQ^2_q}{3s^2}\left(-\frac{t}{s}+
-\frac{s}{t}\right)\label{sigb}
\end{eqnarray}
where $Q_q$ is the quark charge and
\begin{gather}
s=(p_1+p_2)^2,\quad t=(p_1-p_3)^2 \quad{\rm and} \quad u=(p_1-p_4)^2
\label{stu}\\
t=-s(1-\cos \Theta_s)/2 \quad{\rm and} \quad u=-s(1+\cos \Theta_s)/2
\label{tu}
\end{gather}
with the scattering angle $ \Theta_s$ in the di-jet $cms$. 
The  cross section is maximal for small $t$ (or $u$) where it is dominated by
quark exchange.

In the resolved processes the partons ($q,\bar q$ and $g$) in the photon
interact with the partons of the proton.
We restrict ourselves in this discussion to the small angle scattering 
where the gluon exchange contribution dominates and consider  
the subprocesses
\begin{eqnarray}
c)& \qquad\qquad g(p_1)+g(p_2)& \to \;g(p_3)+g(p_4) \label{ggto}\\
d)& \qquad\qquad g(p_1)+q(p_2)& \to \;g(p_3)+q(p_4) \label{gqto}\\
e)& \qquad\qquad q(p_1)+q'(p_2)& \to \;q(p_3)+q'(p_4) \label{qqpto}\\
f)& \qquad\qquad q(p_1)+q(p_2)& \to \;q(p_3)+q(p_4) \label{qqto}\\
g)& \qquad\qquad q(p_1)+\bar q(p_2)& \to \;q(p_3)+\bar q(p_4) \label{qqbarto}
\end{eqnarray}
(in processes $d$ and $f$ the $q$'s could be replaced by the $\bar q$'s).

The different exchanges in the direct and resolved processes cause 
 different distributions $dn/dt\sim t^{-1}$ and $dn/dt\sim t^{-2}$
respectively and this difference is clearly seen in the data \cite{dijzeus}.
Therefore in this case, by choosing suitable intervals of $x_\gamma$, one can
select processes of different colour structure and one can check whether or
in which kinematical interval the low momentum particles are sensitive to
it. To this end we calculate the respective production rates for the soft
particles.

For the given total energy the hard process depends on 3 variables which we
choose here for simplicity as the incoming parton energy fractions $x_1$ and
$x_2$ and the momentum transfer $t$. The cross section for producing the
di-jets with and without accompagnying soft gluons is then given by
\begin{eqnarray}
\frac{d\sigma}{dx_1dx_2dtd\Omega_{\vec n}dp}&=&
   \sum_{ijkl}f_i(x_1,\mu^2)f_j(x_2,\mu^2)\frac{d\sigma^{ij\to kl}}{dt} 
    \frac{\alpha_s }{(2\pi)^2p} W^{ij;kl}(\vec n)
\label{dsigsoft}\\
\frac{d\sigma}{dx_1dx_2dt}&=&
   \sum_{ijkl}f_i(x_1,\mu^2)f_j(x_2,\mu^2)\frac{d\sigma^{ij\to kl}}{dt}    
\label{dsighard}
\end{eqnarray}
where $f_i(x,\mu^2)$ denotes the parton structure functions
at scale $\mu^2$. We
are interested in the density of the soft radiation, i.e. in
the ratio
\begin{equation}
\frac{dN}{d\Omega_{\vec n}dp}=\frac{d\sigma}{dx_1dx_2dtd\Omega_{\vec n}dp}/
    \frac{d\sigma}{dx_1dx_2dt} \label{density}
\end{equation}
where often in practical applications 
an integration over ranges of the kinematical variables
$x_1,x_2$ and $t$ has to be performed. If only one subprocess $[ij\to kl]$
contributes we come back to relations analogous to (\ref{Nqqbg}).\\

\subsubsection{Soft particles in direct processes}
 The angular factors required in (\ref{dsigsoft}) for the partons 
$q$, $\bar q$ and $g$ 
involved in processes a) and b) in
(\ref{gamgl}) and (\ref{gamq})  are given by
(\ref{wqqbg}) with the 
proper relabeling of the partons (see also \cite{lk})
\begin{eqnarray}
W^a(\vec n)&=& N_C
\left[\wh{2}{3}+\wh{2}{4}-\frac{1}{N_C^2}\wh{3}{4}\right], \label{Na}\\
W^b(\vec n)&=& 
N_C
\left[\wh{3}{4}+\wh{2}{4}-\frac{1}{N_C^2}\wh{2}{3}\right]. \label{Nb}
\end{eqnarray}
For the case of perpendicular radiation in the di-jet rest frame we have
simply
\begin{equation}
\wh{2}{3}=\wh{1}{4}= 1+\cos \Theta_s, \quad
\wh{2}{4}=\wh{1}{3}= 1-\cos \Theta_s, \quad
\wh{1}{2}=\wh{3}{4}=2. \label{wh234}
\end{equation}
If either  of the two direct processes a) or b) dominates we obtain for
the respective rates, normalized as in (\ref{rperpdef})
\begin{eqnarray}
R_\perp^a&=&1,\label{rpa}\\
R_\perp^b&=& \frac{N_C}{4C_F}\left[3-\cos \Theta_s-\frac{1}{N_C^2}
   (1+\cos \Theta_s)\right].\label{rpb}
\end{eqnarray}
This means that the soft radiation in the photon gluon fusion process
is the same as in the standard process $\epem \to q\bar q$ 
irrespectively of
the scattering angle $\Theta_s$. In case of the QCD-Compton process we 
find in the limit of small angles $\Theta_s\ll1$ 
which is dominated by quark exchange $R_\perp^b=1$. In the case of
 backward scattering with $\Theta_s\sim \pi$ (i.e. the transition
$\gamma\to g$ at small angle) we have 
effectively a colour octet channel and find $R_\perp^b=N_C/C_F$ as
in case of a $gg$ final state. In the large $N_C$ approximation we can
consider the bremsstrahlung in the quark exchange processes as coming from
a single colour dipole whereby the second is inactive (for $\Theta_s=0$);
 in case of octet exchange the radiation comes from two dipoles and adds  
incoherently. 

If the two jets are not identified one has to add the cross sections
for scatterings at angles $\Theta_s$ and  $\pi - \Theta_s$. One finds
\begin{align}
R_\perp^{a,sym}&=1\label{rasym}\\
R_\perp^{b,sym}&=\frac{N_C}{2C_F}\frac{G(x)\left[1+x-
y/N_C^2\right]+G(y)\left[1+y-x/N_C^2\right]}{G(x)+G(y)}
\label{rbsym}
\end{align}
with $x=-t/s$, $y=-u/s$ and  $G(z)=z+z^{-1}$. So the latter ratio
rises from  $R_\perp^{b,sym}=1$ at  $\Theta_s=0$ to  
$R_\perp^{b,sym}=\frac{13}{8}=1.63$
at  $\Theta_s=\frac{\pi}{2}$.
Therefore, in the kinematic region where one of the processes 
dominates one can check the variation of the soft
particle density with scattering angle $\Theta_s$, analogously to  the case 
of $\epem \to q\bar qg$. In the more general situation one has to use 
the superposition (\ref{dsigsoft}) with the structure functions as weights.

In the further discussion 
we restrict ourselves to the scattering at small
angles where quark or gluon exchanges dominate to 
simplify the discussion.
Then in both processes a) and b)
the quark exchange with $R_\perp^b=1$ dominates and this
result also holds in a general superposition (\ref{dsigsoft})
of symmetrized direct processes. 

\subsubsection{Soft production from resolved processes}
For small angles  $-t\ll s$ the expressions for
the radiation patterns simplify significantly \cite{dkt8}.
The antenna patterns accompagnying the $2\to 2$ parton scattering through
the t-channel exchange are then given by
\begin{eqnarray}
W^c(\vec n) &=&  
N_C \left[\wh{1}{3}+\wh{2}{4}+\frac{1}{2}\left\{
   \wh{1}{2}+\wh{3}{4}+ \wh{1}{4}+\wh{2}{3}\right\}\right], \label{Nc}\\
W^d(\vec n) &=&
N_C \left[\wh{1}{3}-\frac{1}{N_C^2}\wh{2}{4}+\frac{1}{2}\left\{
   \wh{1}{2}+\wh{3}{4}+ \wh{1}{4}+\wh{2}{3}\right\}\right], \label{Nd}\\
W^e(\vec n)&=&W^f(\vec n)\;=\; 2C_F [\wh{1}{4}+\wh{2}{3}  \nonumber\\
&\;&+\frac{1}{2N_CC_F}\left\{
   2(\wh{1}{2}+\wh{3}{4})- \wh{1}{4}-\wh{2}{3}
   - \wh{1}{3}-\wh{2}{4}\right\}], \label{Nef}\\
W^g(\vec n)&=& 2C_F [\wh{1}{2}+\wh{3}{4}  \nonumber\\
&\;&+\frac{1}{2N_CC_F}\left\{
   2(\wh{1}{4}+\wh{2}{3})- \wh{1}{2}-\wh{4}{3}
   - \wh{1}{3}-\wh{2}{4}\right\}], \label{Neg}
\end{eqnarray}
Considering the subprocesses  $i=c, \dots,g$ separately
we obtain the ratios
\begin{eqnarray}
R_\perp^c&=&\frac{N_C}{4C_F}(5-\cos \Theta_s),\label{rpc}\\
R_\perp^d&=& \frac{N_C}{4C_F}\left[4-\frac{1}{N_C^2}(1-\cos \Theta_s)
   \right],\label{rpd}\\
R_\perp^{e,f}&=&(1+\cos \Theta_s)+\frac{1}{N_CC_F},\label{rpe}\\
R_\perp^{g}&=&2+\frac{1}{2N_CC_F}(3\cos \Theta_s -1).\label{rpg}
\end{eqnarray}
At low scattering angles $\Theta_s\ll 1$ 
where the gluon exchange dominates all these ratios $R_\perp^i$ are
\begin{equation}
R_\perp^i=\frac{N_C}{C_F}\approx 2.25\quad (i=c,\dots,f) \label{rpi}
\end{equation}
and for the important processes c) and d) they vary  only slowly with
angle. Therefore also for the superposition (\ref{dsigsoft}) of all processes
the ratio will be close to~(\ref{rpi}) for not too large angles.
On the other hand, these ratios 
differ markedly (by about a factor of 2) from $R_\perp^{a,b}$ for 
the direct
photoproduction. In the large $N_C$ limit and at $ \Theta_s\ll 1$ 
$R_\perp^{c-g}$ correspond to the incoherent sum of two $q\bar q$
antennae-dipoles, while  $R_\perp^{a,b}$ are given by only one  $q\bar q$
antenna.

Therefore, when comparing the perpendicular radiation pattern in direct and
resolved photoproduction of di-jets one expects an increased particle density
in the latter case by about a factor of 2. It will be interesting to find
out experimentally whether such a difference occurs already at such low
momenta as a few hundred MeV. Such a finding would 
provide an important support for the relevance of perturbative QCD,
in particular of the colour coherence, to the
soft particle production.

\subsection{Hard processes in hadron hadron collisions}
The tests which are interesting from our point of view
 are very similar to those discussed for
photoproduction in the last subsection. First there are the subprocesses
(\ref{ggto})-(\ref{qqto}) of parton parton collisions (at small angles)
which we discussed in connection with resolved photoproduction.
Secondly, there are the processes with a final state direct photon
(or, equivalently a weak vector boson W or Z) produced together with a
hadronic jet with opposite transverse momenta (see, e.g. \cite{dkmt1,ks}). 
These processes
correspond to the direct processes in photoproduction.
The relevant subprocesses are
\begin{eqnarray}
a') \qquad q(p_1)+\bar q(p_2)& \to &\gamma(p_3)+ g(p_4), \label{gamglf}\\
b') \qquad q(p_1)+g(p_2)& \to &\gamma(p_3)+\bar q(p_4) \label{gamqf}
\end{eqnarray}
and the soft bremsstrahlung is again found from (\ref{wqqbg})
with appropriate relabeling
\begin{eqnarray}
W^{a'}(\vec n)&=& N_C
\left[\wh{1}{4}+\wh{2}{4}-\frac{1}{N_C^2}\wh{1}{2}\right], \label{Nhb}\\
W^{b'}(\vec n)&=& N_C
\left[\wh{1}{2}+\wh{2}{4}-\frac{1}{N_C^2}\wh{1}{4}\right]. \label{Nha}
\end{eqnarray}
Then we obtain for the perpendicular radiation 
\begin{eqnarray}
R_\perp^{a'}&=&1,\label{rpap}\\
R_\perp^{b'}&=& \frac{N_C}{4C_F}\left[3-\cos \Theta_s-\frac{1}{N_C^2}
   (1+\cos \Theta_s)\right]\label{rpbp}
\end{eqnarray} 
just as in case of direct photoproduction. 

There is an interesting difference with the case of photoproduction. At 
high primary energies the parton collisions with initial gluons are
more likely. Therefore in photoproduction the photon-gluon fusion
process (a) without angular dependence in the soft particle production rate
is favoured whereas at the hadron collider the other process ($b'$) with
angular dependence is more likely. This offers the possibility to study the 
angular dependence of the soft particle production in the same process.
 
\section{Conclusions}

The analytical perturbative approach to multiparticle production together
with the assumption of LPHD has proven to be very successful in the
description of various inclusive characteristics of jets. It is of
importance to investigate further the potential and limitations of this
picture, in particular in the soft region where hadronization effects
could in principle wash out  the perturbative predictions.

The soft particles follow a striking prediction of the
perturbative analysis. The particle density at low momentum is nearly
energy independent over two orders of magnitude in the $cms$ energy 
of $e^+e^-$ annihilation and this behaviour is supported now  
also by the data from 
$ep$ collisions. More generally, the analytic calculations 
suggest a scaling law for particles at low 
transverse momentum; such a property was observed already long
ago in the PETRA energy range. The scaling law is derived analytically
for soft gluons which are emitted coherently from all other more energetic
colour sources. In the perturbative description this process of single gluon
emission is dominated by the Born term of ${\cal O} (\alpha_s)$ which is energy
independent. At larger $p_\perp$ gluons are more frequently produced 
through cascading processes of higher order in $\alpha_s$. 
Remarkably, the hadrons follow this prediction demonstrating the relevance of
LPHD also for soft particles.

It is very important to further clarify whether the observed scaling
behaviour is just a general property of hadronization or should be
considered as consequence of the perturbative QCD as in our picture. 
In the latter case the density of
the low momentum particles should vary in a well defined way with the
type of reactions and orientations of the 
primary partons in the hard process.
In this paper we derive predictions for processes for which data are
already available. Experimental studies of the proposed type would
clarify at which scale of the low transverse momentum or energy the  
behaviour expected from the perturbative analysis sets in.

\section*{Acknowledgements} 

We would like to thank A. De Angelis, Yu.L. Dokshitzer, D. Kant, P.
M\"attig, R. Orava, G. Thompson, T. Sj\"ostrand and E.A. De Wolf for useful
discussions.

\newpage

\section*{Figure captions} 

{\bf Fig.1. a}: Charged particle distribution 
$dn/d^3p$ as a function of particle momentum $p$. 
Experimental data at 
various $cms$ energies\cite{mark1,tasso,topaz,opal1,aleph133} 
are compared with predictions 
using $dn/d^3p = 2\cdot 4/9\cdot K_h D_g^g(\xi_E)/[4 \pi E(E^2 - Q_0^2)]$ 
with $D_g^g$ 
computed in MLLA from eqs.~(\protect\ref{dmlla},\protect\ref{duetermini})
($Q_0$ = 0.27 GeV, $K_h$ = 0.45), which approach a common limit for $p \to$
0. The detailed shape in the dashed region is particularly sensitive 
to parton level approximations. 
{\bf b}: Same distribution for charged 
pions\cite{datapion,topaz,alephrev,opalpion} in comparison 
 with predictions 
using $dn/d^3p = 2\cdot 4/9\cdot K_h D_g^g(\xi_E)/[4 \pi E_h^3]$
with $D_g^g$ computed from the Limiting Spectrum 
($Q_0$ = 0.155 GeV, $K_h$ = 1.125).

\noindent 
{\bf Fig. 2. a}: The same observable as in Fig.~{\bf 1a}, but only the 
data collected at $cms$ energies from
3 up to 58 GeV\cite{mark1,tasso,mark2,topaz} are shown. 
{\bf b}: The same observable as in {\bf a}, but only the 
data collected at LEP at $cms$ energies $\sqrt{s}$ = 91, 133, 161
GeV\cite{opal1,aleph133,alephrev,delphi0,data2b} are shown.

\noindent 
{\bf Fig. 3. a}: Charged particle distribution $dn/d^3p$ at three values
of momentum $p$ = 0.2, 0.7 and 1.2 GeV as a function of $cms$ energy 
$\sqrt{s}$. Experimental data from $e^+e^-$ annihilation 
experiments\cite{mark1,tasso,mark2,topaz,opal1,aleph133,alephrev,%
delphi0,data2b} 
and from $ep$ experiment \cite{H1} in the Breit frame. In the latter case 
two sets of data are
shown, corresponding to different ways of sampling the data (see \cite{H1}); 
in both sets data have been multiplied by a factor 2 to obtain a common
normalization with $e^+e^-$ data and a selection on the energy flow has been
applied~\cite{H1}. The solid line corresponds to formula (\ref{phrel1})
and the dashed one to formula (\ref{phrelLS}). 
{\bf b}: Distribution $dn/d^3p$ of charged kaons and of protons 
at fixed values of momentum as a function of $cms$ energy. 
Experimental data 
from $e^+e^-$ experiments\cite{tassokaon,alephrev,opalpion}.

\noindent 
{\bf Fig. 4}: 
Rapidity density at $y = 1$, $\frac{dn}{dy} \bigl|_{y=1}$, and 
distribution $\frac{1}{5}\frac{dn}{d^3p}$ at momentum $p$ = 0.25 GeV 
for charged particles in $e^+e^-$ annihilation 
 as a function of $cms$ energy $\sqrt{s}$. Data 
 from ~\cite{tassofan,opal1,alephrev,delphi0,data2b}. 

\noindent 
{\bf Fig. 5}: 
 $m^3 dn/d^3p$ distribution for charged pions, charged kaons, protons and
 neutral kaons in $e^+e^-$  annihilation 
as a function of the particle momentum $p$. 
Experimental data from~\cite{datapion,tassokaon,opalpion,alephrev}. 

\noindent 
{\bf Fig. 6}: 
Theoretical prediction at $\sqrt{s}$ = 14 GeV for the  
single inclusive particle density 
$d^2n/dp_{\parallel} d\vartheta$  as a function of the angle
$\vartheta$ at four different values of $p_{\parallel}$ = 0.15, 0.3, 0.5, 0.8
GeV (from right to left) according to MLLA and eq.~\eref{htop} 
 (with parameters $Q_0$ = 0.5 GeV and $\lambda$ = 0.05).

\noindent
{\bf Fig. 7}:   
Theoretical predictions as in Fig.~{\bf 6} for 
$d^2n/dp_{\parallel} dp_\perp$ 
as a function of transverse momentum $p_\perp$ at fixed value of 
$p_\parallel$ = 0.8 GeV  
at four different values of $cms$ energy $\sqrt{s}$ = 14 (solid), 34
(long-dashed), 91 (short-dashed) and 172 GeV (dotted line).

\noindent
{\bf Fig. 8}:   
Comparison of different theoretical predictions 
for the $d^2n/dp_{\parallel} dp_\perp$ distribution 
as a function of transverse momentum $p_\perp$ at fixed value of 
$p_\parallel$ = 0.8 GeV and $\sqrt{s}$ = 34 GeV: DLA Born term
(long-dashed), DLA (short-dashed), MLLA (solid) with $Q_0$ = 0.5 GeV and
$\lambda$ = 0.05 and Limiting Spectrum (dotted line) 
with the same $Q_0$ and $\lambda$ = 0 (normalization adjusted by factor 
$K_h$ = 3.5).

\noindent
{\bf Fig.9}: 
The spectrum of soft particles perpendicular to the event plane depends on
the colour and relative orientation of the primary partons. 

\newpage

\begin{figure}[p]
          \begin{center}
\mbox{\epsfig{file=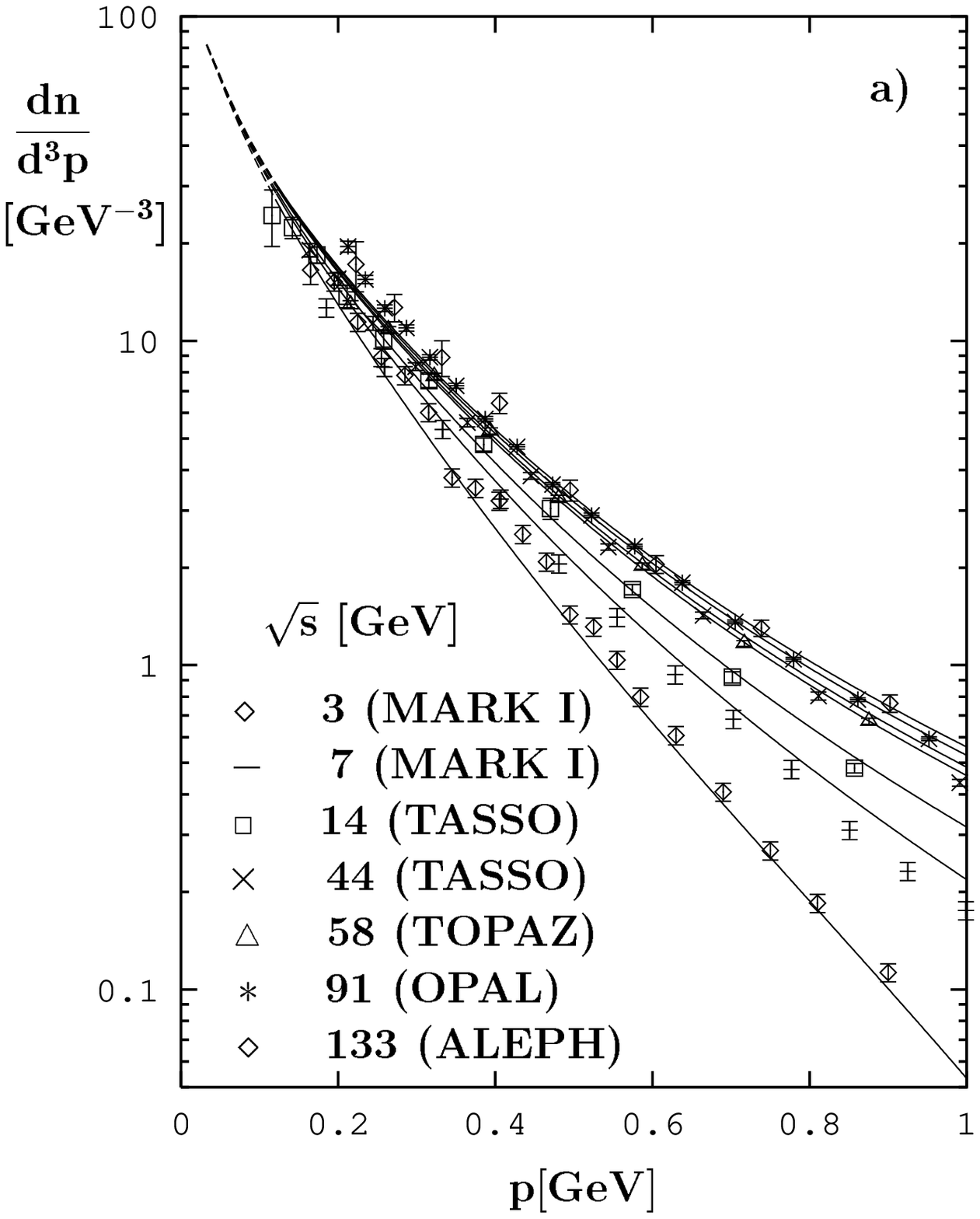,bbllx=4.5cm,bblly=9.5cm,bburx=16.5cm,bbury=26.cm}}
       \end{center}
\caption{}
\end{figure}

\newpage 

\setcounter{figure}{0}

\begin{figure}[p]
          \begin{center}
\mbox{\epsfig{file=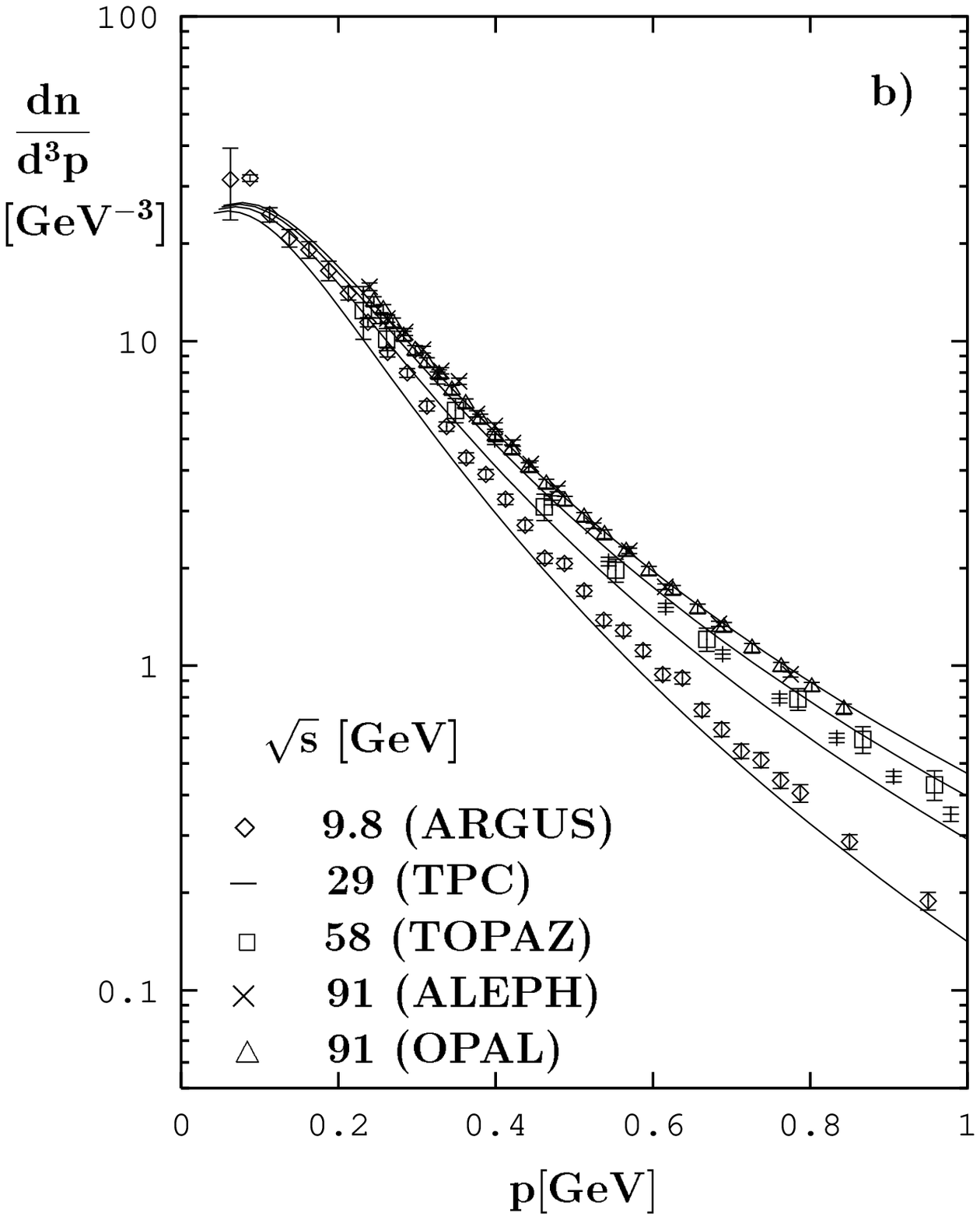,bbllx=4.5cm,bblly=9.5cm,bburx=16.5cm,bbury=26.cm}}
       \end{center}
\caption{}
\end{figure}

\newpage 

\setcounter{figure}{1}

\begin{figure}[p]
          \begin{center}
\mbox{\epsfig{file=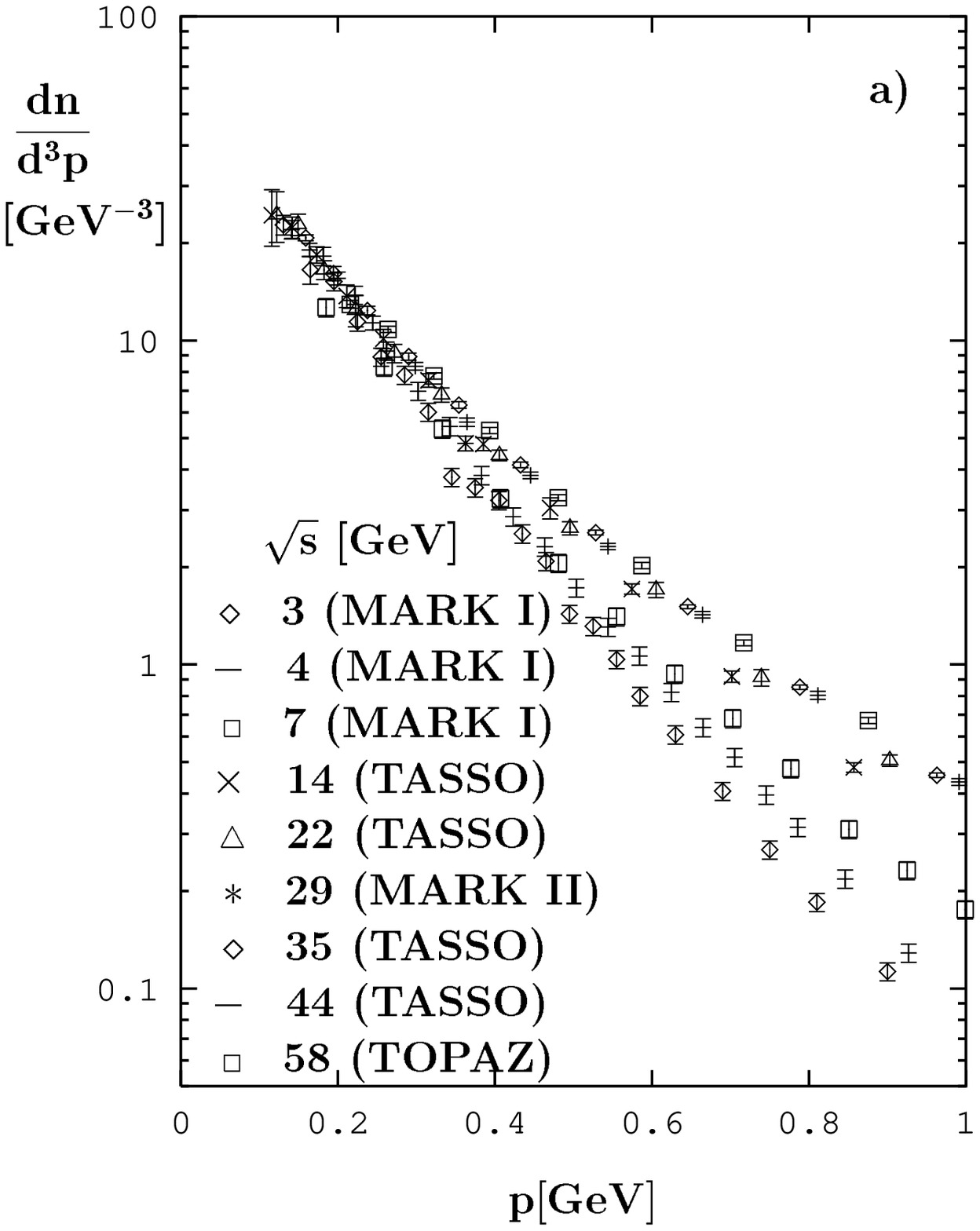,bbllx=4.5cm,bblly=9.5cm,bburx=16.5cm,bbury=26.cm}}
       \end{center}
\caption{}
\end{figure}

\newpage 

\setcounter{figure}{1}

\begin{figure}[p]
          \begin{center}
\mbox{\epsfig{file=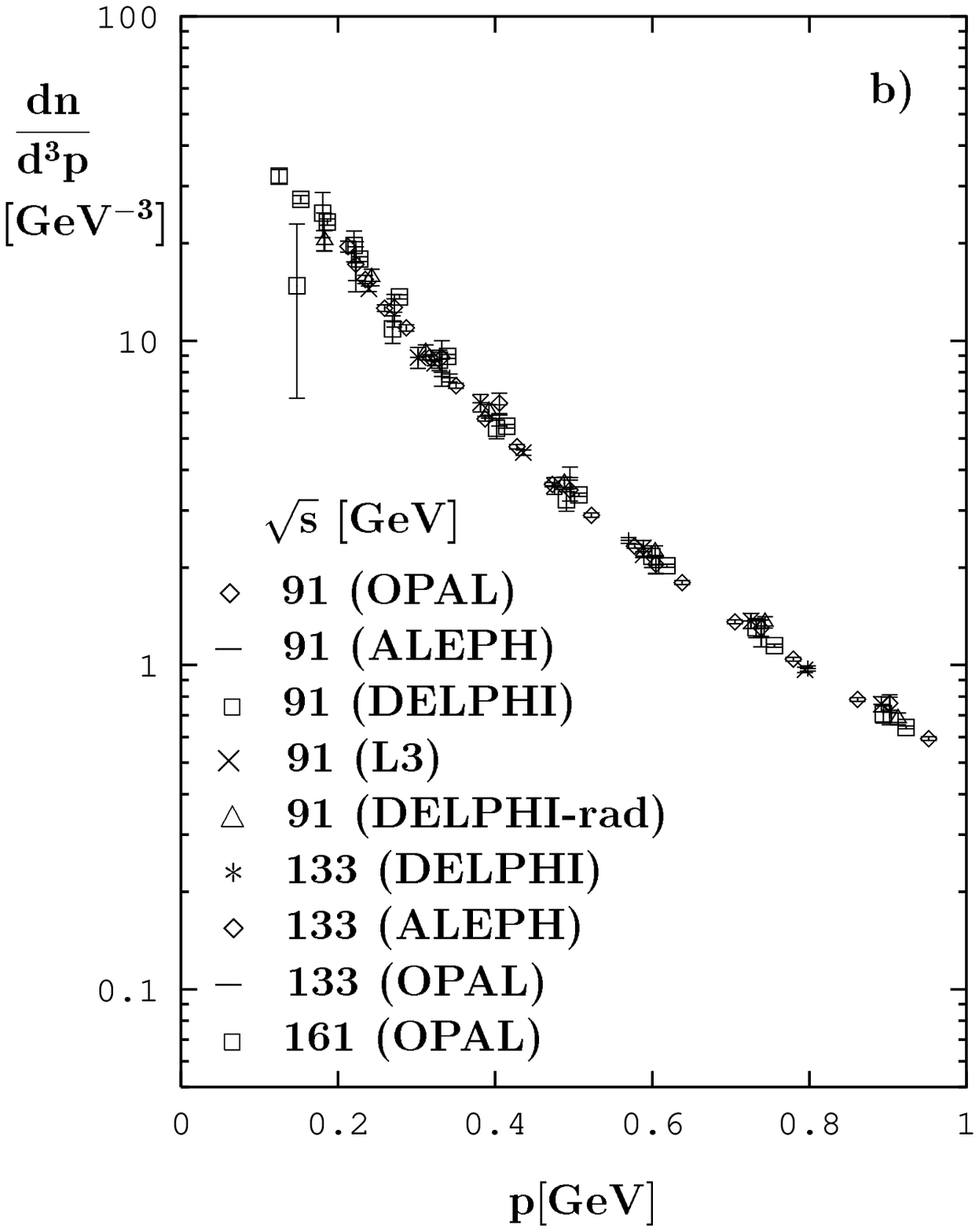,bbllx=4.5cm,bblly=9.5cm,bburx=16.5cm,bbury=26.cm}}
       \end{center}
\caption{}
\end{figure}

\begin{figure}[p]
          \begin{center}
\mbox{\epsfig{file=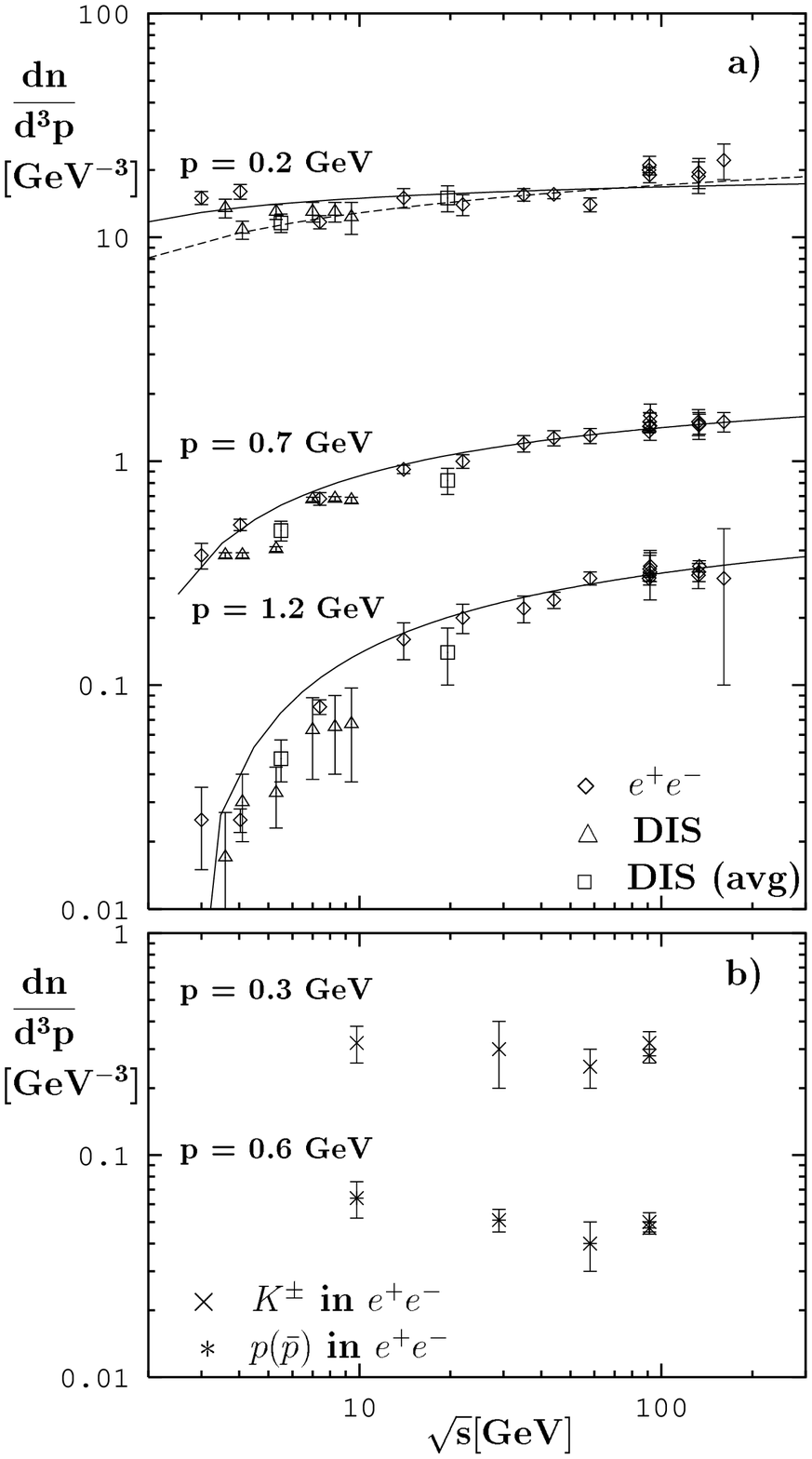,bbllx=4.5cm,bblly=4.5cm,bburx=16.5cm,bbury=26.cm}}
       \end{center}
\caption{}
\end{figure}

\newpage 

\begin{figure}[p]
          \begin{center}
\mbox{\epsfig{file=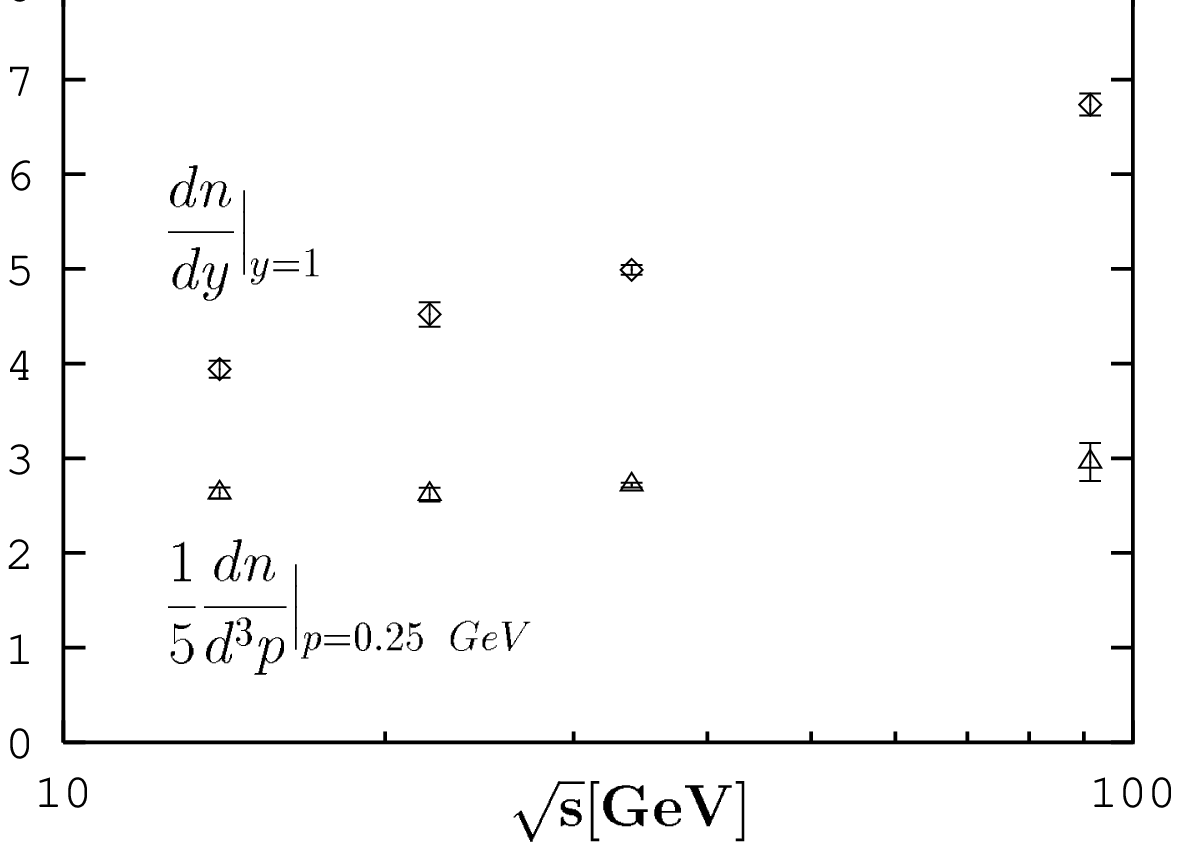,bbllx=4.5cm,bblly=9.5cm,bburx=16.5cm,bbury=26.cm}}
       \end{center}
\caption{}
\end{figure}

\newpage

\begin{figure}[p]
          \begin{center}
\mbox{\epsfig{file=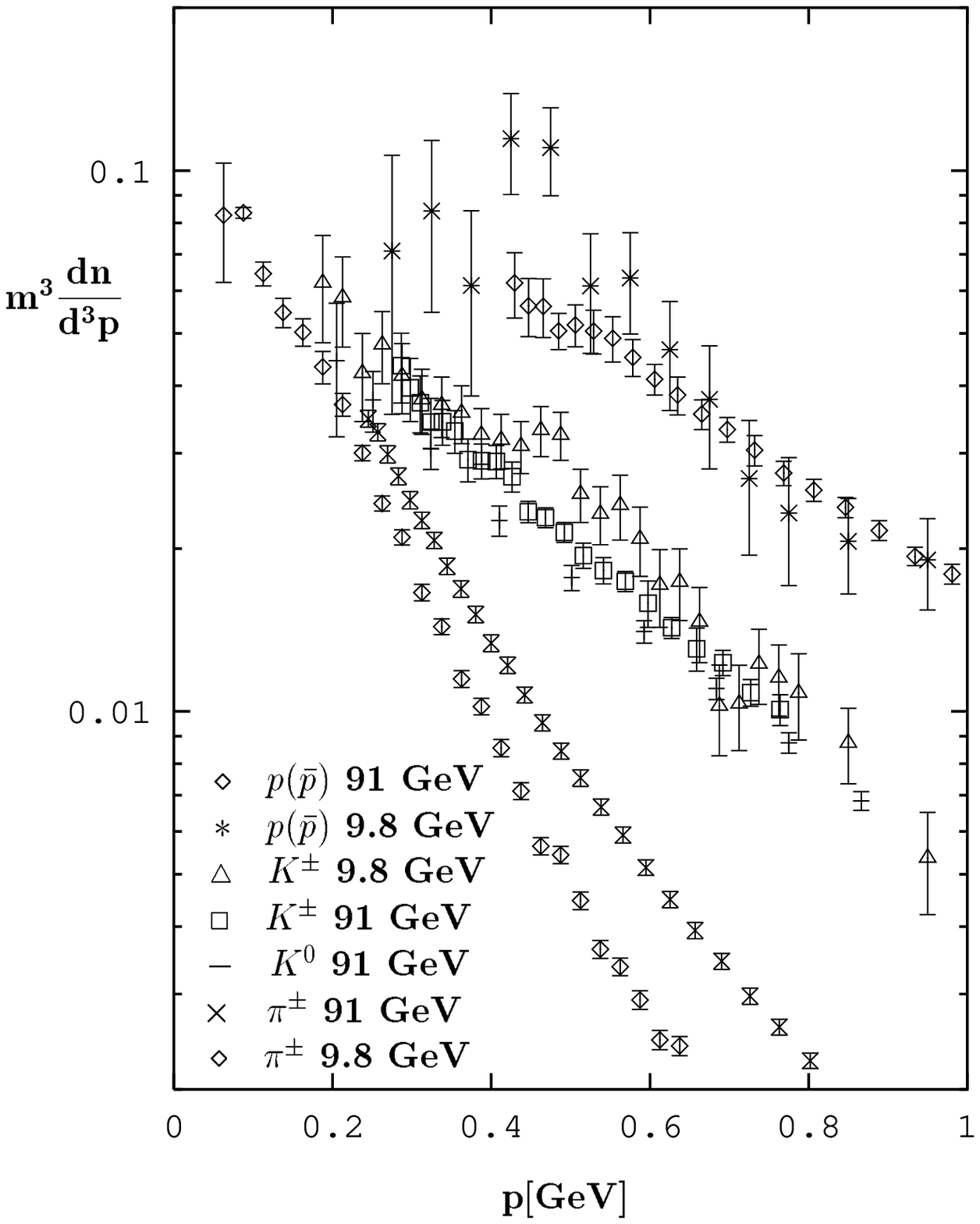,bbllx=4.5cm,bblly=9.5cm,bburx=16.5cm,bbury=26.cm}}
       \end{center}
\caption{}
\end{figure}

\newpage 

\begin{figure}[t]
          \begin{center}
\mbox{\epsfig{file=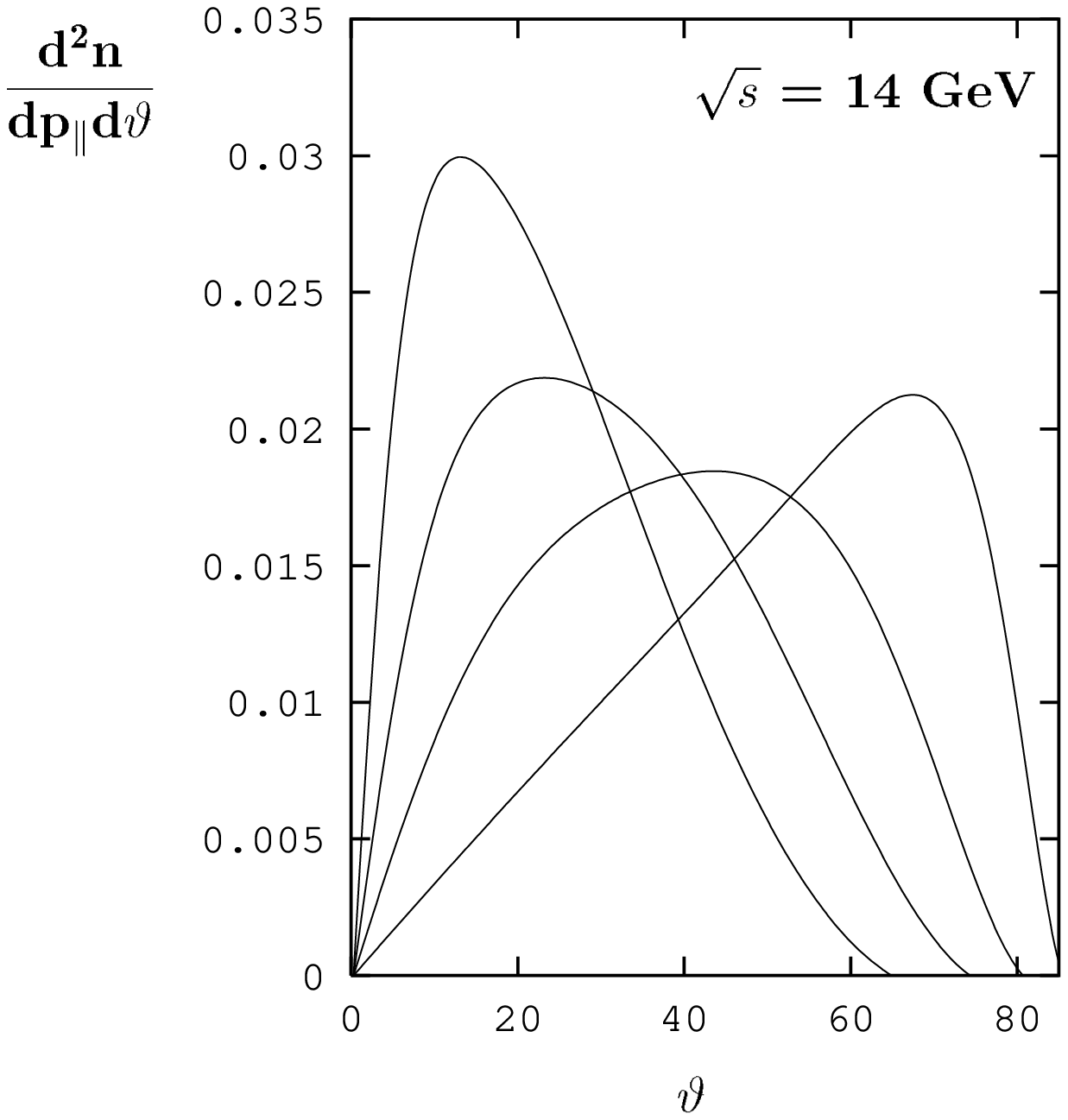,bbllx=5.cm,bblly=11.5cm,bburx=17.cm,bbury=24.cm,width=12cm}}
       \end{center}
\caption{}
\label{figfan}
\end{figure}

\newpage

\begin{figure}[t]
          \begin{center}
\mbox{\epsfig{file=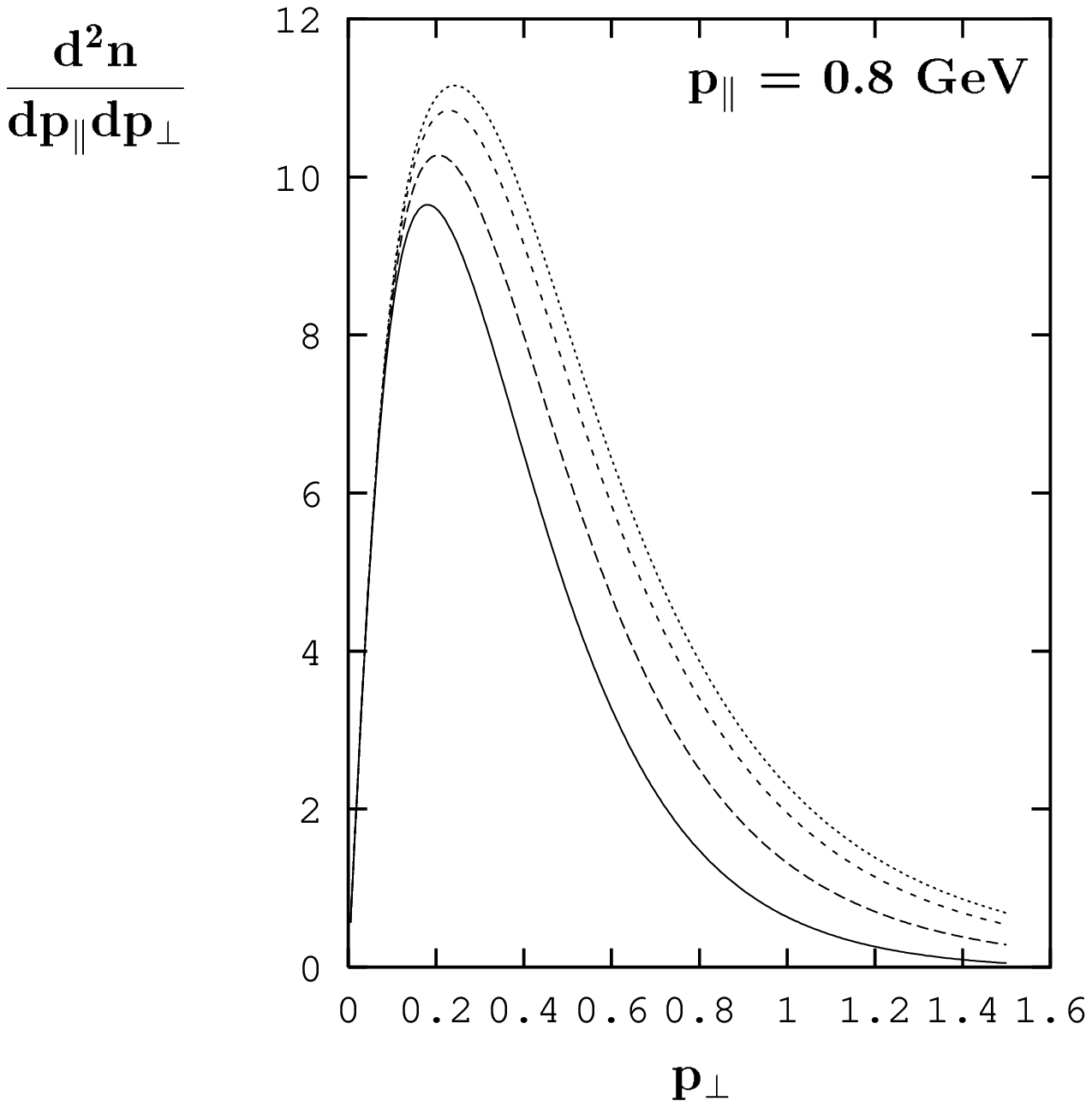,bbllx=5.cm,bblly=11.5cm,bburx=17.cm,bbury=24.cm,width=12cm}}
       \end{center}
\caption{}
\label{energy}
\end{figure}

\newpage

\begin{figure}[t]
          \begin{center}
\mbox{\epsfig{file=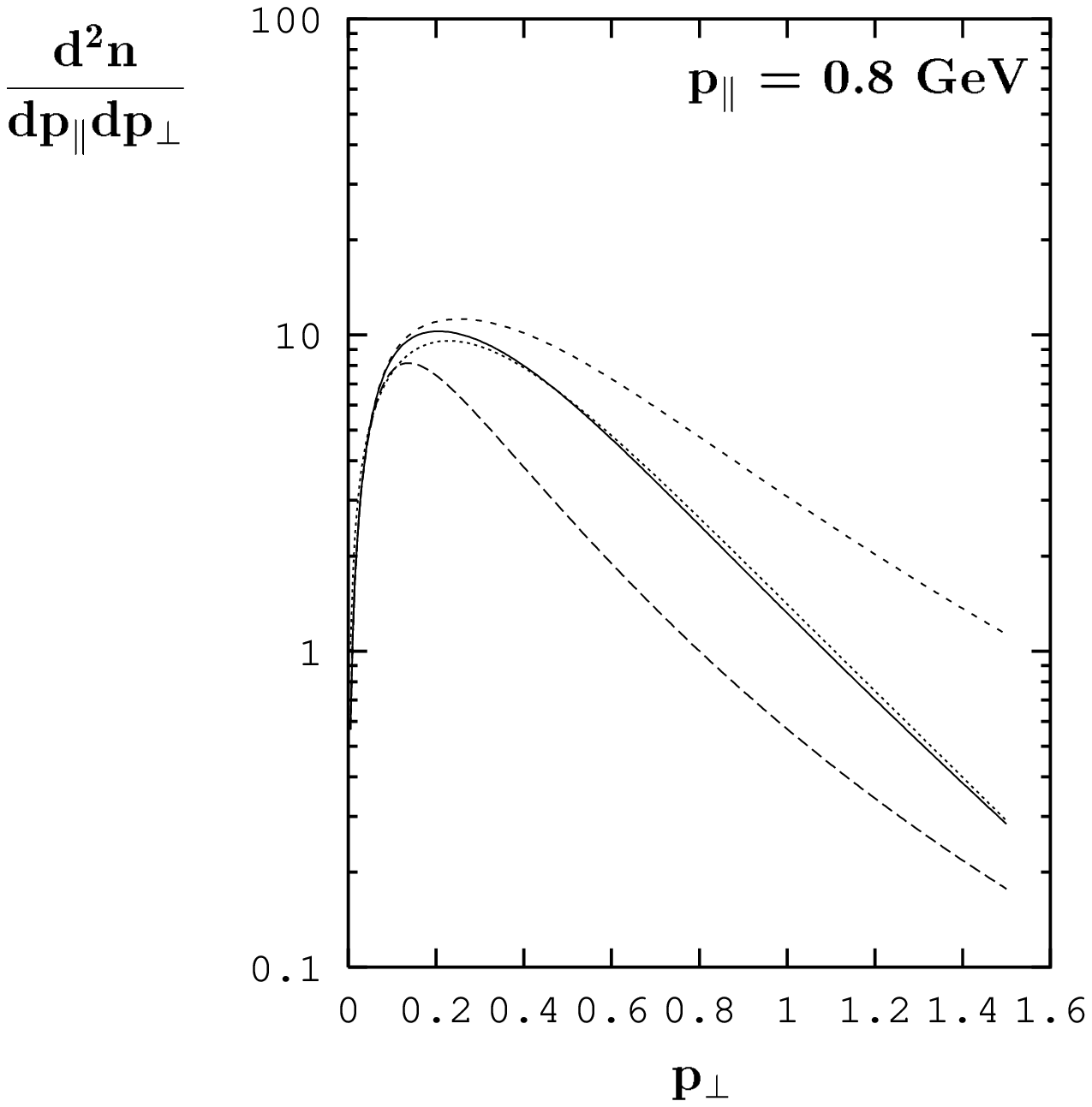,bbllx=5.cm,bblly=11.5cm,bburx=17.cm,bbury=24.cm,width=12cm}}
       \end{center}
\caption{} 
\label{approx}
\vspace{2.0cm}
          \begin{center}
\mbox{\epsfig{file=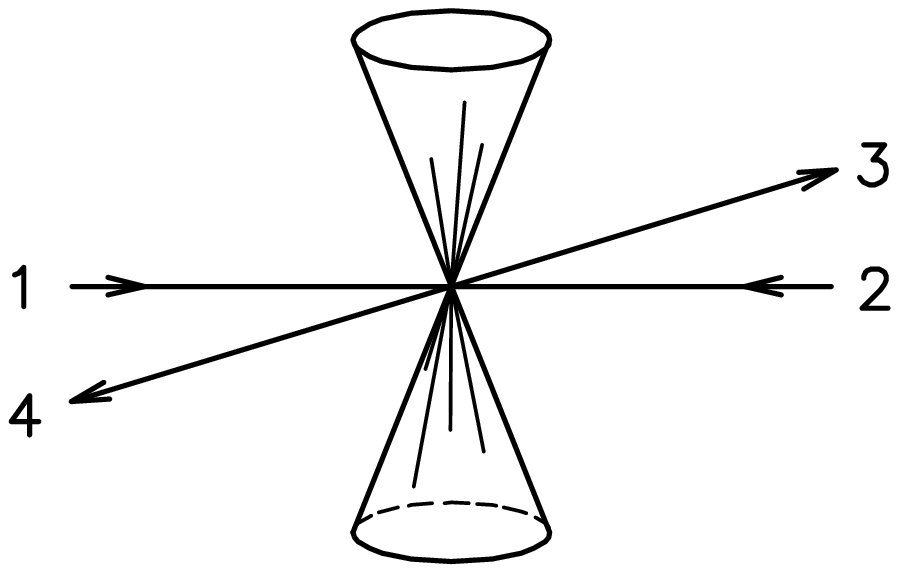,bbllx=3.5cm,bblly=17.cm,bburx=15.5cm,bbury=20.5cm}}
       \end{center}
\caption{}
\end{figure}

\end{document}